\documentclass[12pt]{iopart}
\usepackage[T1]{fontenc}
\usepackage[latin9]{inputenc}
\usepackage{graphicx}
\usepackage{esint}

\makeatletter

\usepackage{iopams}
\usepackage{setstack}

\makeatother

\begin{document}

\newcommand{\mm}[1]{\mathrm{#1}}
\newcommand{\dt}[1]{\frac{\mathrm{d}#1}{\mathrm{d}t}}
\newcommand{\Dt}[1]{\frac{\mathrm{D}#1}{\mathrm{D}t}}
\newcommand{\pp}[2]{\frac{\partial#1}{\partial#2}}
\newcommand{\vek}[1]{\boldsymbol{#1}}
\renewcommand{\vec}[1]{\boldsymbol{#1}}
\renewcommand{\v}[1]{\boldsymbol{#1}}

\title{Memory effects in chaotic advection of inertial particles}

\author{Anton Daitche}

\address{Institute for Theoretical Physics, Münster University, Wilhelm-Klemm-Str.
9, D-48149, Münster, Germany}

\ead{anton.d@wwu.de}

\author{Tamás Tél}

\address{Institute for Theoretical Physics, Eötvös University, and
MTA-ELTE Theoretical Physics Research Group,
P\'azm\'any P. s. 1/A, H-1117 Budapest, Hungary}

\ead{tel@general.elte.hu}

\begin{abstract}
A systematic investigation of the effect of the history force on
particle advection is carried out for both heavy and light
particles. General relations are given to identify parameter regions
where the history force is expected to be comparable with the Stokes
drag.  As an illustrative example, a paradigmatic two-dimensional
flow, the von Kármán flow is taken. For small (but not extremely
small) particles all investigated dynamical properties turn out to
heavily depend on the presence of memory when compared to the
memoryless case: the history force generates a rather non-trivial
dynamics that appears to weaken (but not to suppress) inertial
effects, it enhances the overall contribution of viscosity.  We
explore the parameter space spanned by the particle size and the
density ratio, and find a weaker tendency for accumulation in
attractors and for caustics formation. The Lyapunov exponent of
transients becomes larger with memory. Periodic attractors are found
to have a very slow, $t^{-1/2}$ type convergence towards the
asymptotic form.  We find that the concept of snapshot attractors is
useful to understand this slow convergence: an ensemble of particles
converges exponentially fast towards a snapshot attractor, which
undergoes a slow shift for long times.

\end{abstract}
\maketitle

\section{Introduction\label{sec:Introduction}}

The advection of small, rigid, finite-size particles has been studied
recently in several setups because it plays an important role in many
environment-related phenomena ranging from atmospheric science and
oceanography to astrophysics (for reviews see
\cite{Magnaudet2000,Toschi2009,Cartwright2010}). For example, inertial
particles are important in cloud
microphysics~\cite{Falkovich2002,Grabowski2013}, planet
formation~\cite{Bracco1999}, and aggregation and fragmentation
processes in fluid flows~\cite{Nishikawa2001,Zahnow2009,Zahnow2011}.
A particularly interesting aspect is the advection of such particles
in chaotic flows where experimental results
\cite{Ouellette2008,Sapsis2011} are available by now. Applications can
be for example the forecasting of pollutant transport for homeland
defense \cite{Benczik2002,Nishikawa2001} and the localization of a
toxin or biological pathogen spill (e.g. anthrax) from outbreaks in a
street canyon \cite{Tang2009}.

Due to the drag acting on the particles, the Hamiltonian passive
advection problem is converted into a dissipative problem which can
have attractors.  Correspondingly, inertial particles can have the
tendency to accumulate in certain regions of the flow \cite{Bec2007}
(a phenomenon termed preferential concentration). A characteristic
feature in comparison with the dynamics of ideal tracers is the
appearance of caustics \cite{Falkovich2002,Wilkinson2005}, i.e., the
intersection of different branches of the chaotic sets in the
configuration space. This is a consequence of the fact that what we
see is a projection of the full pattern of the high-dimensional phase
space to the configuration space of the fluid. The main relevance of
the existence of caustics is that the probability of collisions of
particles becomes enhanced in such regions of the flow.

The basic equation of motion for small spherical particles in a
viscous fluid describable in the Stokes regime is given by the
Maxey-Riley equations \cite{Maxey1983,Gatignol1983}\footnote{It is
  interesting to note that Gatignol derived the same equation as Maxey
  and Riley in the same year. Currently, it is this equation augmented
  with the corrections introduced by Auton and coworkers
  \cite{Auton1988} (see equation~(\ref{eq:MR-Equations-dimensional})) that
  is commonly called the ``Maxey-Riley equation''}. Its precise form
contains an integral, also called the history force, which describes
the diffusion of vorticity around the particle during its full time
history. The kernel of this integral is therefore proportional to the
power -1/2 of the elapsed time.

The origin of such a kernel can perhaps best be seen in the problem of
an infinitely large plate moving in a rather viscous fluid.  If the
plate starts at time zero with a sudden change in the velocity which
is then kept constant, in Stokes approximation, one solves the
diffusion equation for the velocity field.  The solution contains a
prefactor $1/\sqrt{t}$. For the case of a general motion of the plate,
one imagines to break up the problem in a set of jumps in the plate's
velocity at different times $\tau$. Each such jump contributes a
factor to the flow velocity at time $t$ proportional to
$1/\sqrt{t-\tau}$ multiplied by the velocity jump. In a continuous
time picture, the result for the shear stress on the plate is
proportional to an integral from zero to $t$ of the plate's
acceleration at time $\tau$ multiplied by the kernel $1/\sqrt{t-\tau}$
(see Landau-Lifschitz \cite{Landau1987}, paragraph 24).  For the
motion of a spherical particle in a viscous flow it were Boussinesq and
Basset (at the end of the 19th century) who pointed out the appearance
of such a history force (for historical details see
\cite{Michaelides2006}).  This force is therefore sometimes called the
Boussinesq-Basset or the Basset force (we use the term history force).
The equation of motion of a moving sphere in a fluid at rest was
already given in Landau-Lifschitz (\cite{Landau1987}, problem 7 to
paragraph 24).  The precise form of the equation in a moving fluid
\cite{Maxey1983,Gatignol1983,Auton1988}, i.e., equation~(\ref{eq:MR-Equations-dimensional}) or
(\ref{eq:MR-Equations-dimensionless}),  was, however, established only
a century after the first works of Boussinesq and Basset.

The history force renders the advection
equation of inertial particles to be an integro-differential equation.
Because of this difficulty, the memory represented by this integral
term was neglected in the majority of papers on chaotic advection
of inertial particles.

In non-chaotic flows, the history force was shown to have important
effects \cite{DO1994,Mordant2000,Abbad2004,Hill2005, Angilella2007}. A
particularly interesting example is an experiment
\cite{Lohse2006,Lohse2009} which points out the determining importance
of memory in the motion of microbubbles propelled by ultrasound. In
this setting the trajectories, and a surprising destabilization of
these (in spite of rather strong dissipation), can be understood only
when the history force is taken into account. A very recent experiment
with suspended particles indicates the importance of the history force
also in the phenomenon of Brownian motion \cite{Kheifets2014}. Within
a perturbative treatment restricted to the limit of very weak inertia,
it has been found that the history force can suppress chaotic
advection \cite{YRK1997}.  In contrast to this, memory has recently
been found to be able to enhance chaos in particle advection in the
presence of gravity when settling or rising takes place
\cite{Guseva2013}.  In turbulent flows, the effects of the history
force have been studied in
\cite{Reeks1984,Mei1991,Armenio2001,Aartrijk2010,Hinsber2011,Calzavarini2012,Olivieri2014},
showing that this force can have significant influence. For neutrally
buoyant particles it has been suggested in \cite{Sapsis2011} that
there are dynamical effects beyond the history force, Faxén
corrections and the lift force.

Our aim is to carry out a comprehensive investigation of long term
memory effects due to strong inertia in smooth flows with chaotic advection, made possible by a
recent progress in the numerical treatment of the problem
\cite{Daitche2013}. Concerning the physical aspects, we are extending
here the results of a short paper \cite{Daitche2011} before which, to our knowledge,
no such investigations have been carried out.

The paper is organized as follows. First, we review the basic equation
(section~\ref{sec:Basic-equations}), and provide estimates for
parameters where memory effects are expected to be relevant
(section~\ref{sec:estimates}). Next we sketch the applied numerical
scheme (section~\ref{sec:numscheme}). In section~\ref{sec:Model-flow},
a paradigmatic two-dimensional chaotic model flow is introduced, that
of the von K\'arm\'an vortex street, which shall be used as an
illustrative example throughout the paper. The choice of the particle
parameters is discussed in section~\ref{sec:parameters}. In chaotic
flows, the basic question is not so much about the deviation between
trajectories with and without memory (treated briefly in section
\ref{sec:Trajectories}), rather the deviation in \emph{statistical}
properties. We shall point out in section~\ref{sec:escape-dynamics}
that properties like the escape rate or the lifetime statistics of
particle ensembles significantly differ due to the history force both for
bubbles and aerosols (particles lighter and heavier than the fluid).
The residence time distribution and the value of the power-law
exponent of non-hyperbolic decay is also different in the two cases, as
shown in section~\ref{sec:residence-times}. Other types of statistics,
like e.g. that of the different forces acting on particles while being
in the wake is investigated in section~\ref{sec:PDFs}.  One important
effect of memory is that the tendency for accumulation is much weaker
than with Stokes drag only and, as a consequence, attractors occur
only at rather special parameters in this open flow.  When attractors
nevertheless exist, the temporal convergence of individual
trajectories towards them is converted from an exponential approach to
a power-law approach (section~\ref{sec:attractors}).  It is worth,
however, replacing the traditional individual trajectory picture by a
global one, and monitoring the evolution of a particle ensemble. In
this picture an exponential convergence is found to a so-called
snapshot attractor, which then slowly moves towards the traditional
attractor (section~\ref{sec:snapshot-attractors}).  After briefly
discussing also cases with chaotic attractors
(section~\ref{sec:chaoticattr}), we show in section~\ref{sec:saddle}
that the average Lyapunov exponents of transient chaos with memory are
larger than without.  The details of how to define and determine
Lyapunov exponents in an integro-differential equation, like the
Maxey-Riley equation, are relegated to \ref{sec:appendix-lyapunov}.
The paper is concluded with a summary of several dynamical features
illustrating our main finding: memory effects change the inertial
dynamics so that it moves towards that of the passive dynamics, but in
a highly non-trivial way.  We also return to the discussion of the
relevance of the history force and argue that the size parameter is
most naturally defined in terms of the shortest characteristic
time scale. A refined estimate is found, valid for any flow, enabling
us to also briefly discuss the relevance of memory effects in
turbulent flows.

\section{Basic equations\label{sec:Basic-equations}}

The equation of motion of a rigid spherical particle of radius $a$ and
mass $m_{p}$ in a fluid of kinematic viscosity $\nu$, density
$\rho_{f}$ and velocity field $\vek u(\vek r,t)$ reads (including
gravity) as \cite{Maxey1983,Gatignol1983,Auton1988}
\begin{eqnarray}
m_{p}\dt{\vek v} & = & m_{f}\Dt{\vek u}-\frac{m_{f}}{2}\left(\dt{\vek v}-\Dt{\vek u}\right)-6\pi a\rho_{f}\nu\left(\vek v-\vek u\right) + (m_p-m_f) {\vec g} \label{eq:MR-Equations-dimensional}\\
 &  & -6a^{2}\rho_{f}\sqrt{\pi\nu}\int_{t_{0}}^{t}\mathrm{d}\tau\frac{1}{\sqrt{t-\tau}}\left(\frac{\mathrm{d}\vek v}{\mathrm{d}\tau}-\frac{\mathrm{d}\vek u}{\mathrm{d}\tau}\right).\nonumber
\end{eqnarray}
Here $\vek v\equiv \mm d\vek r/\mm d t$ is the particle velocity, $m_{f}$ is
the mass of the fluid excluded by the particle, ${\vec g}$ the
gravitational acceleration, and
\begin{equation}
\dt{\vek u}=\pp{\vek u}t+\vek v\cdot\nabla\vek u
\end{equation}
and
\begin{equation}
\Dt{\vek u}=\pp{\vek u}t+\vek u\cdot\nabla\vek u
\end{equation}
denote the full derivative along the trajectory of the particle and of
the corresponding fluid element, respectively.  The terms on the
right-hand side of (\ref{eq:MR-Equations-dimensional}) are: the force
exerted by the fluid on a fluid element at the location of the
particle, the added mass term\footnote{We note that in the original
  derivation by Maxey and Riley, and by Gatignol the added mass term
  contained the derivative $\dt{\vek u}$. Later it was found that
  $\Dt{\vek u}$ is a better choice as it is correct in a wider range
  of conditions (see \cite{Auton1988,Maxey1993,Babiano2000}).}  describing the impulsive pressure response of
the fluid, the Stokes drag, the buoyancy-reduced weight, and the
history force. The history force, an integral term, accounts for the
viscous diffusion of vorticity from the surface of the particle along
its trajectory \cite{Maxey1983}.

Equation~(\ref{eq:MR-Equations-dimensional}) is valid if the initial
particle velocity $\vek v(t_{0})$ at time $t_0$ matches the fluid
velocity. Otherwise, an initial-velocity-dependent term
\cite{Michaelides1992,Maxey1993},

\begin{equation}
-6a^{2}\rho_{f}\sqrt{\pi\nu}\frac{\vek v(t_{0})-\vek u(t_{0})}{\sqrt{t-t_{0}}}\label{eq:initialv}
\end{equation}
should be added to the right-hand side of (\ref{eq:MR-Equations-dimensional}).
By means of the identity (that can be derived by partial integration)
\begin{equation}
\frac{1}{\sqrt{\pi}}\int_{t_{0}}^{t}\frac{\frac{\mathrm{d}}{\mathrm{d}\tau}\left(\vek v-\vek u\right)}{\sqrt{t-\tau}}\mathrm{d}\tau+\frac{1}{\sqrt{\pi}}\frac{\vek v(t_{0})-\vek u(t_{0})}{\sqrt{t-t_{0}}}=\frac{1}{\sqrt{\pi}} \dt{}\int_{t_{0}}^{t}\frac{\vek v-\vek u}{\sqrt{t-\tau}}\mathrm{d}\tau
\label{eq:memory_derivative}
\end{equation}
we arrive at the most general form of the history force (the
right-hand side of (\ref{eq:memory_derivative})) valid for {\em any}
initial condition. Note that this is formally nothing but
$\left(\frac{\mathrm{d}}{\mathrm{d}t}\right)^{1/2}(\vek v-\vek u)$,
the fractional derivative of order $1/2$ of the slip velocity $\vek
v-\vek u$ (more precisely it is a fractional derivative of the
Riemann-Liouville type \cite{Podlubny1998}). A new feature of the
right-hand side of (\ref{eq:memory_derivative}) is that the nominator
of the integrand does no longer contain a derivative. It is just the
velocity difference what appears there, a feature that makes the
numerical evaluation of the integral easier.  The advantage of using
identity (\ref{eq:memory_derivative}) was pointed out in
\cite{Daitche2013}.

Using relation (\ref{eq:memory_derivative}) and measuring time and
velocity in units of $T$ and $U$, the dimensionless Maxey-Riley
equation valid for any initial condition becomes
\begin{equation}
\frac{1}{R}\dt{\vek v}=\frac{3}{2}\Dt{\vek u}-\frac{1}{S}\left(\vek v-\vek u-W \vek n\right)-\sqrt{\frac{9}{2\pi}\frac{1}{S}}\dt{}\int_{t_{0}}^{t} \frac{\vek v-\vek u}{\sqrt{t-\tau}} \mathrm{d}\tau,\label{eq:MR-Equations-dimensionless}
\end{equation}
where $\vec n$ is a unit vector pointing upwards (against gravity).
Here three dimensionless parameters appear, the density parameter
\begin{equation}
R=\frac{2m_{f}}{m_{f}+2m_{p}}=\frac{2 \rho_{f}}{\rho_{f}+2\rho_{p}},
\label{eq:R}
\end{equation}
where $\rho_p$ denotes the particle density, the size parameter
\begin{equation}
S=\frac{2}{9}\frac{a^{2}/\nu}{T},\label{eq:inertp}
\end{equation}
a ratio of the particle's viscous relaxation time and the characteristic
time $T$ of the flow, and
the dimensionless settling velocity
\begin{equation}
W=S \left(\frac{3}{2}-\frac{1}{R}\right) \frac{gT}{U}.\label{eq:w}
\end{equation}
This is negative (positive) for heavy (light) particles, called
aerosols (bubbles).  We shall see that the size parameter $S$ is a
more appropriate number to characterize memory effects than the
traditional Stokes number, which is $St=S/R$ in our notation.

An important condition for the validity of equation~(\ref{eq:MR-Equations-dimensionless})
is that the particle Reynolds number
\begin{equation}
Re_{p}=\frac{\left|\vek v-\vek u\right|a}{\nu}\label{eq:pReynolds}
\end{equation}
remains small during the entire dynamics \cite{Maxey1983}. In
addition, $S$ must be small (i.e., the particle's characteristic
time scale is much smaller than that of the flow) and the particle
radius must be much smaller than the characteristic linear scale $L$
of the flow: $a\ll L$. The last condition assures that the so-called
Faxén corrections are negligible \cite{Maxey1983}.

In a recent manuscript Farazmand and Haller \cite{Farazmand2013} prove
the local existence of the solutions to
(\ref{eq:MR-Equations-dimensionless}). With initial particle velocity
$\vek v(t_{0})$ matching the fluid velocity, the solution is twice
continuously differentiable in time, but otherwise it might be only
differentiable for $t>t_0$.

Several attempts have been made to extend the Maxey-Riley equation to
finite particle Reynolds numbers (up to a few hundreds) by modifying
the particular form of the forces
\cite{Lovalenti1993,Mei1994,Dorgan2007}. Part of all of these
approaches is a different form of the history force and in some cases
also an empirical nonlinear expression for the drag (see
\cite{Loth2009} for a review).  Concerning the range of applicability
of the Maxey-Riley equation, it is interesting to note that Maxey and
Wang \cite{Maxey1996} found that up to $Re_{p}\approx17$ even equation
(\ref{eq:MR-Equations-dimensional}) ``may be quite adequate in
practice even though not justified by theory''. In an experimental
study \cite{Abbad2004} considering particle Reynolds numbers up to
$0.5$, the standard form of the history force has been found to remain
valid. In our simulations the average particle Reynolds number is on
the order of unity (see section~\ref{sec:PDFs}). Therefore, we expect the
standard Maxey-Riley equation to be adequate for the advection of
inertial particles in our case and decide not to consider any finite
Reynolds number corrections to the Maxey-Riley equation. In this study
we focus on the changes induced by the history force in the motion of
inertial particles. Therefore we do not consider effects like e.g.,
the lift-force.


Equation~(\ref{eq:MR-Equations-dimensionless}) is a second-order
integro-differential equation for the trajectory $\vek r(t)$ of an
inertial particle. To any initial condition $(\vek r_{0},\vek v_{0})$
at time $t_{0}$ there is a unique trajectory, but the transition
between infinitesimally close neighboring time instants $t$ and
$t+\mm d t$ does not only depend on the state $\vek q(t)\equiv(\vek r(t),\vek v(t))$,
but also on all the previous instants, since the kernel of the integral
decays rather slowly. Considering a stroboscopic map taken at integer
multiples of some time unit, we obtain a map whose value at time $n+1$
depends on its value at all previous integers back to the initial time instant:

\begin{equation}
\vek q_{n+1}=\vek M_{n}(\vek q_{n},\vek q_{n-1},...,\vek q_{1,}\vek{q_{0}}).\label{eq:strobm}
\end{equation}
From a dynamical systems point of view, the dynamics is thus that
of a non-autonomous map with memory whose range is increasing as time
goes on. This is a non-standard problem in which novel features can
show up. It is of particular interest which kind of chaos and attractors can
be present in such systems.

\section{Estimating the relevance of the history force} \label{sec:estimates}

To estimate when the history force is important, we compare its magnitude to the
Stokes drag (both are viscous effects). From the dimensionless Maxey-Riley equation
(\ref{eq:MR-Equations-dimensionless}), the ratio of these two forces is seen to scale with
$\sqrt{S}$. Denoting the typical magnitude of the history force and the drag
by $F_h$ and $F_d$ respectively, we can write
\begin{equation}
\frac{F_h}{F_d}=\sqrt{\frac{9}{2}}\alpha\sqrt{S} ,
\label{eq:ratio}
\end{equation}
where $\alpha$ represents the ratio
\begin{equation}
\alpha= \frac{\left|\left(\frac{\mathrm{d}}{\mathrm{d}t}\right)^{1/2}(\vek v -
  \vek u)\right|}{|\vek v - \vek u|} =
\frac{\left|\frac{1}{\sqrt{\pi}}\frac{\mathrm{d}}{\mathrm{d}t}\int_{t_0}^t
  \frac{1}{\sqrt{t-\tau}} (\vek v - \vek
  u)\mathrm{d}\tau\right|}{|\vek v - \vek u |}.
\label{eq:beta}
\end{equation}
The value of $\alpha$ might depend on the process in question and is
expected to be about unity (see also the discussion in section \ref{sec:discussion}).
Equation~(\ref{eq:ratio}) provides an estimate for the magnitude of the
history force relative to the Stokes drag.

For practical purposes we might consider the history force to be
negligible if its magnitude is smaller then $1\%$ of the Stokes drag,
i.e.,  if $S<10^{-4}({2/9})/\alpha^2$. It might appear that $1\%$ is
already quite small. However, note that we are making order of
magnitude estimates here and could be easily off by a factor of
$10$. In terms of the particle radius $a$ in (\ref{eq:inertp}) this
$1\%$-condition means
\begin{equation}
a <\frac{1}{100 \alpha}\sqrt{\nu T}.
\label{eq:a}
\end{equation}
In order to relate this condition to the Reynolds number $Re=UL/\nu$
of the flow, we define the Strouhal number $Sl$ as the ratio of $L/U$
and the characteristic time $T$ of the flow
\begin{equation}
Sl =\frac{L}{UT}
\label{eq:Sl}
\end{equation}
and find
\begin{equation}
\frac{a}{L}<\frac{1}{100 \alpha \sqrt{Sl Re}}
\label{eq:aL}
\end{equation}
as a condition under which memory effects can be neglected.  Note that
the above argument assumes a smooth flow. For multi-scale flows
(turbulence) the length scale $L$ and the parameters $Re$ and $Sl$ in
(\ref{eq:aL}) should be characteristic of the small scales (see also the
discussion in section~\ref{sec:discussion}).

For example, in a smooth flow with $Re=100$ and $Sl \approx 1$ the particle
size has to be about $1000 \alpha$ times smaller then the
characteristic length scale of the flow to be able to neglect the
history force with some confidence.  For $100\,\mathrm{\mu m}$ particles in a
case with $\alpha \approx 1$ this is {\em not} fulfilled in small-scale
flows with $L<10\, \mathrm{cm}$.

Another approach is to consider the weakly inertial limit, where $S$
is assumed to be a small expansion parameter.
Following Druzhinin and
Ostrovsky \cite{DO1994} we obtain the following expression for the
velocity of a nearly ideal particle:
\begin{equation}
\hspace*{-2.3cm}
\dt{\vec r}=\vek u +W \vec n +\frac{S}{R}\left(\frac{3}{2}R-1\right)\left\{ \Dt{\vek u}-\sqrt{S}\sqrt{\frac{9}{2\pi}}\dt{}\int_{t_{0}}^{t}\mathrm{d}\tau\,\frac{1}{\sqrt{t-\tau}}\Dt{\vek u}\right\} +\mathcal{O}(S^{2}).\label{eq:expansion}
\end{equation}
This is a first-order equation for the trajectory and we see that
there are two correction terms to the passive tracer dynamics: the
fluid acceleration and the history force based on the fluid acceleration. Both
corrections are proportional to $S/R$. This ratio determines thus the
importance of inertial corrections.  Provided that $\mm D{\vek u}/\mm
D t$ and the derivative of the integral within the parenthesis are of
the same order, the importance of memory is determined by $\sqrt{S}$,
as also found in the previous consideration (when weak inertia was not
assumed).  Therefore, estimates (\ref{eq:a}), (\ref{eq:aL}) also hold
in the weakly inertial limit (when, of course, the definition of
$\alpha$ should be adjusted according to (\ref{eq:expansion})).

Note that the above considerations do not depend on the density of the
particle. 
Thus, the weight of the {\em history force} is set by the size
parameter $S$, independent of the particle's density. The density
parameter $R$, however, does influence the importance of {\em inertial
  effects}, as $S/R$ is the prefactor of the correction in
(\ref{eq:expansion}).  The natural parameter to estimate the
importance of inertial effects is thus the Stokes number $S/R\equiv
St$, and the size parameter $S$ is the natural quantity to
estimate the importance of the history force.

We would like to say a few words about the case of very heavy
particles, e.g. water droplets in air, $R\approx10^{-3}$.  All our
previous considerations apply to this case as well.  A new feature of
heavy aerosols is, however, that inertial effects might be important
in spite of the smallness of $S$ since $R$ is also small. In such cases
the pressure term $3R/2 \; \mm D\vec u/\mm D t$ in
(\ref{eq:MR-Equations-dimensionless}) can always be neglected, but not
the others.  The relative weight of the history force remains given by
$ \alpha\sqrt{S}\sqrt{9/2}$ and since $S$ is small, memory effects are
not expected to be very important. For $S \sim R \approx 10^{-3}$, $\alpha
\approx 1$, for example, the effect is estimated to be $7\%$ of the
Stokes drag.  When $S <10^{-4}$,
\begin{equation}
  \frac{\mathrm{d}\vek v}{\mathrm{d}t}=-\frac{R}{S}\left(\vek v-\vek u -W \vec n\right)\label{eq:drag-only}
\end{equation}
is an (at least) $2\%$-accurate advection equation for heavy aerosols,
often used in the literature, see
e.g. \cite{Falkovich2004,Bec2007,Saw2012}.  For even smaller heavy
particles: $S\!<\!\!<\!R$, the equation simplifies to the passive
advection equation with settling: $\vek v=\vek u +W \vec n$.

Finally, we mention that the requirement of small particle Reynolds number
(\ref{eq:pReynolds}) provides another inequality for the particle size.
For memory effects to be relevant, inequality (\ref{eq:aL}) should hold with
a reversed sign and this sets a {\em lower} bound to $a/L$.
The conditions that $Re_p$ should not become too large provides an {\em upper} bound
on the dimensionless particle size.

\section{Numerical scheme}
\label{sec:numscheme}

The history force poses the main difficulty for a numerical integration
of (\ref{eq:MR-Equations-dimensionless}). The first major problem
is the singularity of the kernel $1/\sqrt{t-\tau}$ at the upper end of the integration,
at $\tau=t$. Although the singularity is integrable and the history
force is well defined, the numerical evaluation of the kernel near
it leads to large errors when standard quadrature schemes are used.
Recently, a custom quadrature scheme has been developed \cite{Daitche2013}
 where the
kernel is treated analytically. It is expressed as a weighted sum
\begin{equation}
\int_{t_{0}}^{t}\frac{\vek v-\vek u}{\sqrt{t-\tau}}\,\mm d\tau\approx\sum_{i}\mu_{i}\left(\vek v(\tau_{i})-\vek u(\tau_{i})\right),
\end{equation}
where coefficients $\mu_{i}$ contain the effect of the kernel.  The
derivation and specification of the coefficients for higher orders is
quite involved \cite{Daitche2013}. This quadrature scheme is then
incorporated in a linear multistep method of the Adams-Bashforth
type, where relation (\ref{eq:memory_derivative}) is used to evaluate
an integral of the history force.  The quadrature scheme for the
history integral, as well as the whole integration scheme for the
Maxey-Riley equation, has been tested in cases where analytical
solutions are available, see \cite{Daitche2013}, and the order of
convergence has been measured and verified in these cases.  The
numerical results shown here are obtained with a third-order scheme,
i.e., the one-step error is proportional to $(\Delta t)^{4}$ where
$\Delta t$ is the time step.

The second major problem is the high computational costs for a
numerical integration, stemming from the necessity to recompute the
history force --- an integral over all previous times steps --- for
every new time step. Therefore the computational costs grow with the
square of the number of time steps and can become quite substantial
for long integration periods. For the integration of particle
ensembles we have addressed this difficulty by parallel computation
(on ca. $100$ CPUs).

\section{Model flow\label{sec:Model-flow}}

An accurate evaluation of the history term is computationally rather
intensive, and we would like to use particle ensembles of about
$10^{6}$ particles in order to have reliable statistics. Therefore,
we choose an analytically given model flow to limit the numerical
efforts. As a paradigmatic example, we take the kinematic model of a
rather typical fluid phenomenon, the time-periodic shedding of von
Kármán vortices in the wake of a cylinder as described in
\cite{Jung1993} (sometimes called the JTZ flow). This analytical
two-dimensional flow was shown to faithfully represent the
Navier-Stokes dynamics at a (radius-based) Reynolds number $Re\approx125$
and has since then been used in several studies of passive
\cite{Ziemniak1994,Pentek1995,Toroczkai1998,Sandulescu2006,Sandulescu2007,Sandulescu2008}
and inertial \cite{Benczik2002,Benczik2003,Haller2008b,Haller2008}
chaotic advection in environmental flows. In the JTZ model two
vortices are present in the wake at any time (a generalization to more
than two vortices has been given recently in \cite{Wu2010}). The flow
is from left to right.

The period $T$ of vortex shedding is traditionally chosen as the time
unit, and the radius of the cylinder as the length unit $L$.  The
velocity of the vortex cores is on the order of $L/T$, whereas
the inflow velocity is a factor $14$ times larger and is taken as the
velocity unit $U$.  Thus the Strouhal number (\ref{eq:Sl}) is
$Sl=1/14$\footnote{The Strouhal number in the von K\'arm\'an problem
  is often used in its diameter-based form which corresponds then to
  $1/7$.}. One important further parameter is the dimensionless
strength of the vortices $w=192/\pi$ (for comparison with
\cite{Benczik2003} we shall also use the value $w=24$).  The
stream-function of the model is taken from the literature
\cite{Benczik2003}.  Since the flow is two-dimensional, we do not
consider here the effect of gravity and hence $W=0$ in
(\ref{eq:MR-Equations-dimensionless}).  For a recent investigation of
the effect of the history force on sedimentation of aerosols and on
rising of bubbles we refer to \cite{Guseva2013}.

Throughout the simulations presented in this paper the initial
velocity difference between particle and fluid is zero, i.e., $\vek
v(t_{0})-\vek u(t_{0},\vek r_{0})=\vek 0$.  Furthermore, most
simulations are performed with the starting time
$t_{0}=0.2$ (time zero corresponds to the instant when a vortex is
born along the upper right edge of the cylinder). This value has been
chosen to minimize the collisions between particles and the
cylinder. As the fraction of colliding particles is reasonably small,
we have avoided the inclusion of an extra collision rule on the
surface and decided to stop integration in such cases and considered
the particle to be escaped.

When studying the dynamics of inertial particles without memory, a
stroboscopic map can be defined, taken at time instants which are
integer multiples of the period $T$ of the flow. This is an autonomous
map, in which periodic orbits and invariant sets are defined as usual.
In the presence of the memory term, the stroboscopic map becomes of
the form of (\ref{eq:strobm}), in which periodicity after a finite
amount of time is hardly possible, since the state of any time instant
is influenced by the entire history. It is therefore a kind of surprise
that periodic attractors might exist, as will be seen later.

\section{Choice of particle parameters}\label{sec:parameters}

The density parameter (\ref{eq:R}) ranges from $0$ to $2$. Throughout
the paper we shall use two illustrative values $R=0.4$ and $R=2$
typical for aerosols and bubbles, respectively. The former corresponds
to a density ratio $\rho_p/\rho_f=2$ and can be considered to represent
sand grains in water. Since air is about thousand times lighter than
water, the value of $R=2$ can be seen to correspond to the case of spherical
air bubbles in water.

For completeness, we mention that water droplets in air are described
by a density parameter $R=1.2 \times 10^{-3}$. This is an example of
the case of heavy aerosols discussed briefly in
section \ref{sec:estimates}. For oil droplets (bubbles) in water we
obtain with the density $\rho_p= 0.9$ g$/$cm$^3$ of oil $R=0.7$. This
is close to $R=2/3$ of neutrally buoyant particles.

Next we point out that the instantaneous particle Reynolds number
(\ref{eq:pReynolds}) can be determined from the data of our model
flow. Replacing $a$ and $\nu$ in (\ref{eq:pReynolds}) by using (\ref{eq:inertp}) and
$\nu=LU/Re$ we obtain
\begin{equation}
Re_{p}=\sqrt{\frac{9}{2} S \frac{Re}{Sl}}\left|\vek v-\vek u\right|,\label{eq:Re-slip-relation}
\end{equation}
where $\left|\vek v-\vek u\right|$ is the dimensionless slip velocity
(measured in units of $U$).  This will be used to check the validity
of the basic equation~(\ref{eq:MR-Equations-dimensionless}).

The value of the size parameter $S$ will be changed in the range $[0.005,0.09]$,
whereas $Re$ and $Sl$ are set by the flow (to 125 and $1/14$, respectively).

\section{Sample trajectories\label{sec:Trajectories}}

\begin{figure}
\noindent
\centering{}\includegraphics[width=7.7cm]{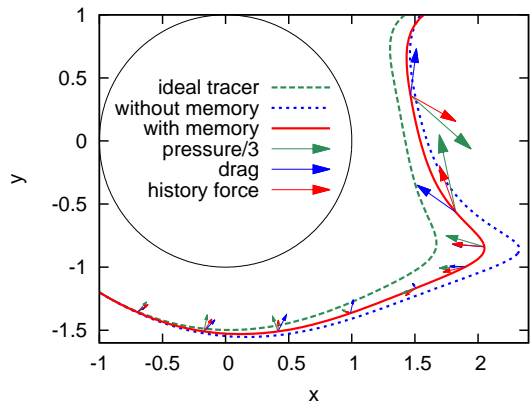}
\caption{
  \label{fig:sample-trajectories-aerosols}Trajectory
  of a single heavy particle ($R=0.4$ - sand in water, $S=0.01$) with respect to different dynamics: with
  memory (red), without memory (blue), and as an ideal tracer (green);
  initiated at $t_{0}=0.2$, $\vek
  r_{0}=\left(-1,\,-1.2\right)$ with vanishing slip velocity.  The
  forces acting on the particle in the presence of memory are shown as
  arrows of different colors (red being the history force). The pressure term is presented with only
  a third of its magnitude. }
\end{figure}

First, we consider trajectories starting with the same initial
condition, and compare different
dynamics. Figure~\ref{fig:sample-trajectories-aerosols} shows the
inertial trajectory of an aerosol with and without memory along with
the trajectory of an ideal tracer (which follows the fluid velocity
exactly). One immediately observes that the trajectories differ and
that (at the beginning) the trajectory with memory is in-between the
other two.

The acceleration of the particle can be decomposed into the three
contributions on the right-hand side of
(\ref{eq:MR-Equations-dimensionless}).  These terms, called the
pressure term, the Stokes drag and the history force are plotted at
subsequent time instants in
figure~\ref{fig:sample-trajectories-aerosols}.  As the caption
indicates, one third of the pressure term is given, for better
visibility. We thus clearly see that the pressure term always
dominates the other two, but the history force is of the same order as
the Stokes drag. A comparison of the full trajectory with that
obtained without the memory term (red and blue lines in
figure~\ref{fig:sample-trajectories-aerosols}, respectively) shows that the direction of
separation (or approach) of theses trajectories coincides with the
direction of the history force.
For bubbles, similar findings have been reported in
\cite{Daitche2011}.

We also carried out simulations with equation of motion
(\ref{eq:expansion}) valid in the weakly inertial limit.  The results
(not shown in figure~\ref{fig:sample-trajectories-aerosols}) indicate a
strong deviation from the true trajectory (red line), as strong as the
deviation between this curve and any other curves in
figure~\ref{fig:sample-trajectories-aerosols}. This shows that a value
as small as $S=0.01$ of the size parameter is not yet small enough to
ensure the weakly inertial limit.  At the same time, this is a sign
indicating that dynamics (\ref{eq:expansion}) does not define a slow
manifold \cite{Haller2008b,Haller2008} for this $S$ (although it is
expected to do so for sufficiently small $S$ values).

\section{Ensemble and escape dynamics\label{sec:escape-dynamics}}

\begin{figure}
\begin{centering}
\includegraphics[width=0.7\columnwidth]{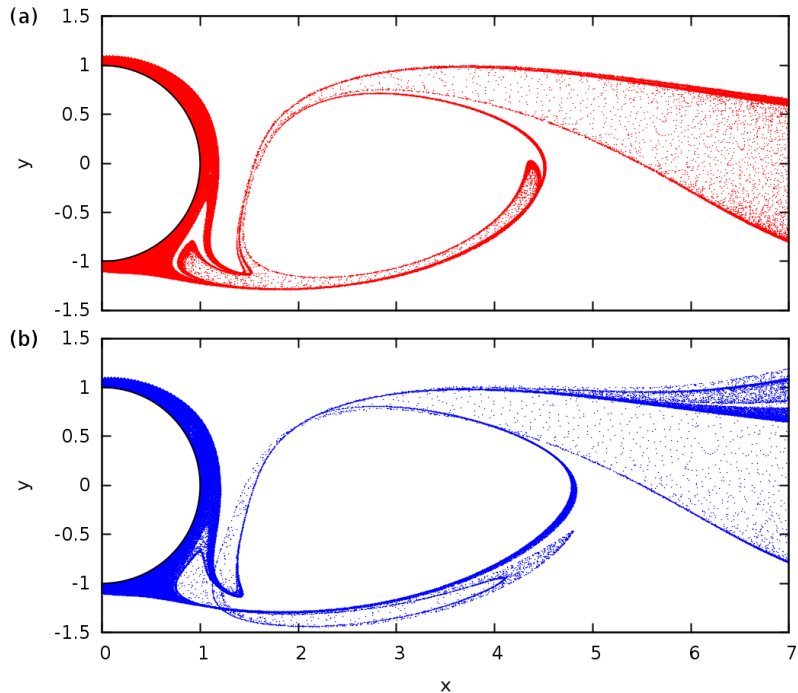}
\par\end{centering}

\caption{\label{fig:ensemble-aerosols}Dynamics of an ensemble of
  $N_{0}\approx1.8\cdot10^{6}$ aerosols ($R=0.4$ - sand in water,
  $S=0.01$) with (a) and without (b) memory starting in the domain
  $\left[-1,4\right]\times\left[-2,2\right]$ around the cylinder at
  $t_{0}=0.2$. The distribution of the particles is shown at time
  $t=0.94$.}
\end{figure}

Next, we turn to the dynamics of particle ensembles. A large number
$N_{0}\approx1.8\cdot10^{6}$ of particles is distributed uniformly in
the domain $\left[-1,4\right]\times\left[-2,2\right]$ around the
cylinder at $t_{0}=0.2$. All particles are followed up to a certain
time when we plot their position. The results for aerosols are shown
in figure~\ref{fig:ensemble-aerosols}. The two distributions differ both
in large- and small-scale structures.  A characteristic feature of the
case without the history force are caustics \cite{Falkovich2002, Wilkinson2005},
i.e., the intersection of different branches in the configuration
space. This is due to the fact that what we see is a projection of the
full pattern in the four-dimensional $(x,y,v_{x,}v_{y})$ phase space
to a plane. Figure~\ref{fig:ensemble-aerosols} shows that the
frequency of occurrence of caustics seems to be \emph{lower} with
memory. We have found similar results for other values of $R$: $0.3$,
$0.5$, $1.0$, $1.2$, $1.5$, $1.7$, $2.0$. This effect of the history
force is even stronger for bubbles \cite{Daitche2011}.  A quantity
closely related to caustics is the collision rate. In accordance to
the suppression of caustics, the history force has been found in
\cite{Daitche2011} to lead to a reduction of the collision rate.
A recent simulation \cite{Olivieri2014} indicates a related phenomenon
in turbulence: the history force is found to reduce preferential
concentration (an effect which enhances collision rates).

In figure~\ref{fig:ensemble-aerosols} a filamentary pattern can be seen
both with and without the history force. This is in itself an
indication for the chaoticity of the advection dynamics in both cases.
Since the problem is dissipative, chaotic attractors might be present.
Irrespective of their existence,
chaos is always present in the form of
transient chaos with an underlying chaotic set, a chaotic saddle
\cite{Ott2002,Tel2006}.

In the transient case all particles eventually escape downstream and
the decay dynamics is best followed by monitoring the total number
$N(t)$ of particles not yet escaped a given region up to time t.  We
distribute $N_{0}\approx1.5\cdot10^{6}$ particles uniformly in the
domain $\left[0.6,4\right]\times\left[-2,2\right]$ outside the
cylinder at initial time $t_{0}=0.2$. A particle is considered to have
escaped if it crosses the line $x=5$ or it enters a circle of radius
$r=1.014$ around the cylinder. The using of a circle larger in radius
than the cylinder is motivated by excluding very slow (non-hyperbolic)
decay characterizing the boundary layer, as done in
\cite{Benczik2002,Benczik2003}. This definition of ``escape'' will be
used throughout the paper, unless stated otherwise (as, e.g., in
section \ref{sec:residence-times}). The number $N(t)$ of survivals is
then determined as a function of time $t$. We find monotonous
exponential decays with different characteristics for different cases.

\begin{figure}
\noindent \begin{centering}
\includegraphics[width=6.05cm]{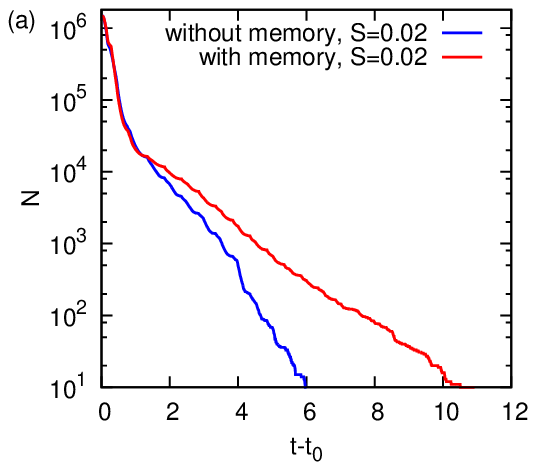}\includegraphics[width=6.05cm]{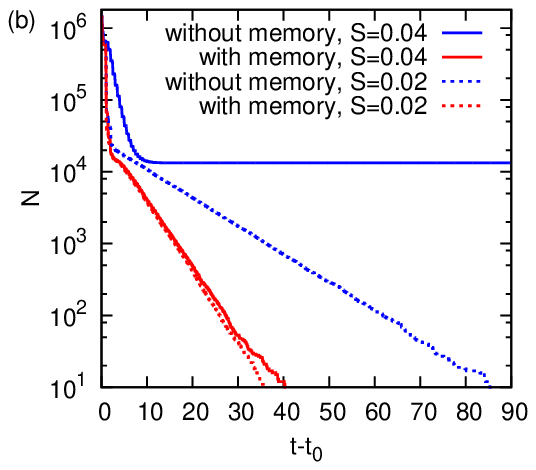}
\par\end{centering}
\caption{\label{fig:Nt} The number $N(t)$ of non-escaped particles for (a) $R=0.4$ - sand in water
(b) $R=2.0$ -  air bubble in water, with two different size parameters. }
\end{figure}

When the particles are heavier than the fluid (aerosols) typically all
of them eventually escape; an example is shown in
figure~\ref{fig:Nt}a. We see that the emptying process is slower with
memory for aerosols.  For bubbles attractors can appear for certain
parameter values. An attractor presents itself as a plateau in $N(t)$,
i.e., a fraction of all the particles never leaves the wake of the
cylinder. See for example the case $S=0.04$ without memory in
figure~\ref{fig:Nt}b; with memory the plateau, and thus the attractor,
disappears. Here we see a profound change in the dynamics of inertial
particles induced by the history force: the disappearance of
attractors. When there are no attractors in the bubble case, the
history force leads to a faster emptying of the wake. In the case
of $S=0.02$ of figure~\ref{fig:Nt}b the time for a complete emptying
roughly halves when memory is included. Note that the influence of the
history force on the speed of the emptying process is opposite for
aerosols and bubbles: memory slows down the emptying for aerosols and
speeds it up for bubbles.

\begin{figure}
\noindent \begin{centering}
\includegraphics[width=0.5\columnwidth]{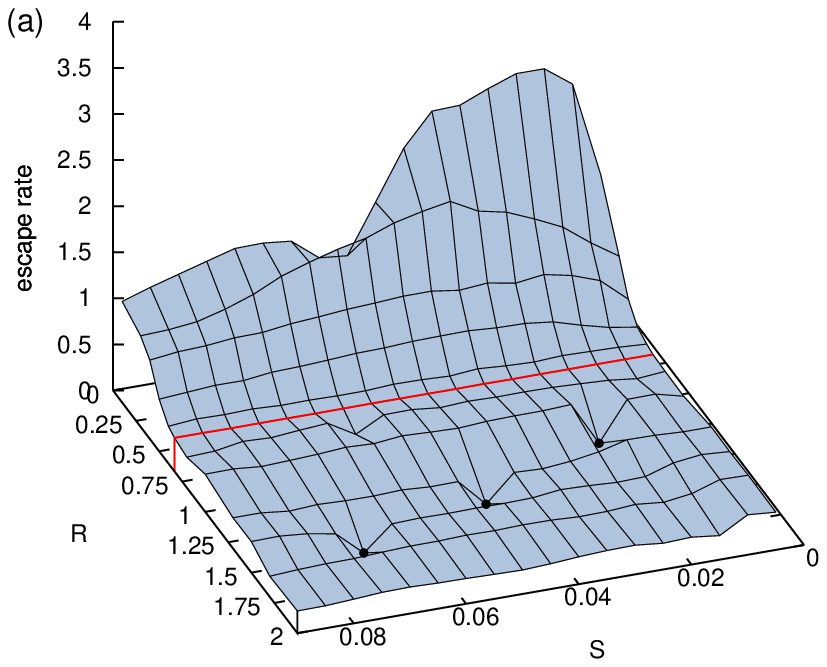}\includegraphics[width=0.5\columnwidth]{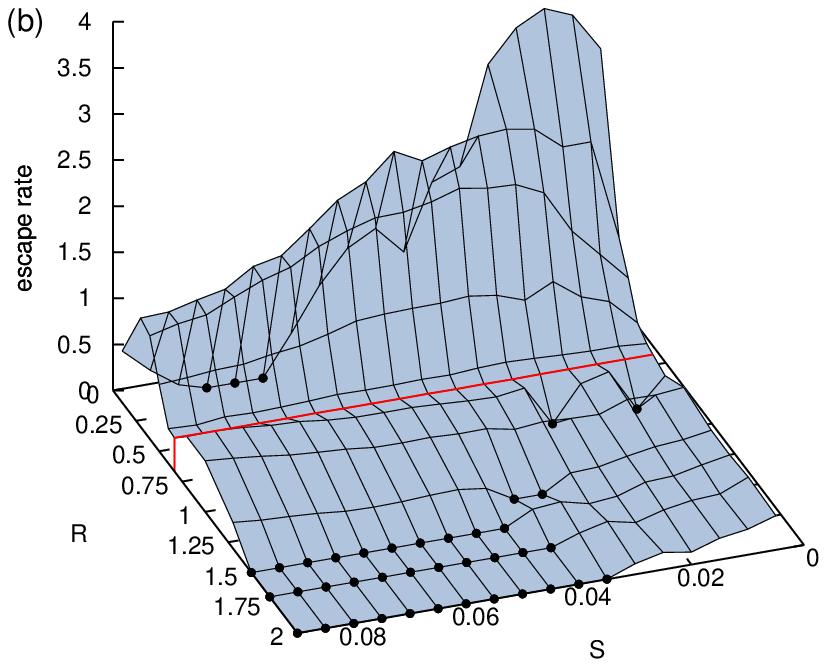}
\par\end{centering}

\caption{\label{fig:escaperates-S}Escape rates on the parameter plane
  of size parameter and density ratio: $(S,R)$ with (a) and without
  (b) memory.  The black dots mark parameters where an attractor is
  present. The $R$-values used as a grid of the plot are 0.1, 0.3,
  0.4, 0.5, 0.6, $2/3$, 0.8, 1.0, 1.3, 1.5, 1.7, 2.0. The red line
  highlights the case of $R=2/3$, i.e., neutrally buoyant particles.}
\end{figure}

A quantitative measure characterizing the escape dynamics from the
wake is the escape rate \cite{Ott2002,Tel2006}. It is the rate $\kappa$
of the exponential decay
\begin{equation}
N(t)\sim\exp(-\kappa t),
\end{equation}
which sets in after an initial transient time, see figure~\ref{fig:Nt}.  In
the case of a plateau in $N(t)$ (due to the presence of an attractor)
we set the escape rate formally to zero. Figure~\ref{fig:escaperates-S}
shows the escape rate with and without memory as a function of the
parameters $R$ and $S$. The red line marks the case of a neutrally
buoyant particles ($R=2/3$), where the escape rate does not depend on
$S$; indeed it is equal to the one of ideal tracers
($\kappa_{\mm{ideal}}=0.364$).  We did not find any influence of memory
on the behavior of neutrally buoyant particles. They behave as ideal
tracers in this flow when the initial slip velocity is zero.

Attractors are quite ubiquitous without memory, see black dots in 
figure~\ref{fig:escaperates-S}b.  One even finds attractors in the
aerosol range $R<2/3$. Since heavy particles are pushed outside of
closed orbits due to the centrifugal force, and intend therefore to
escape, attractors are expected to be atypical for aerosols. In
special cases they were pointed out to exist~\cite{Vilela2007, Angilella2014} and our
results provide further examples of such cases.
A striking effect of the history force is that it kills attractors at
the parameters where they exist without memory. Only very few
attractors (none for aerosols) are found with memory.

\begin{figure}
\noindent \begin{centering}
\includegraphics[width=8.8cm]{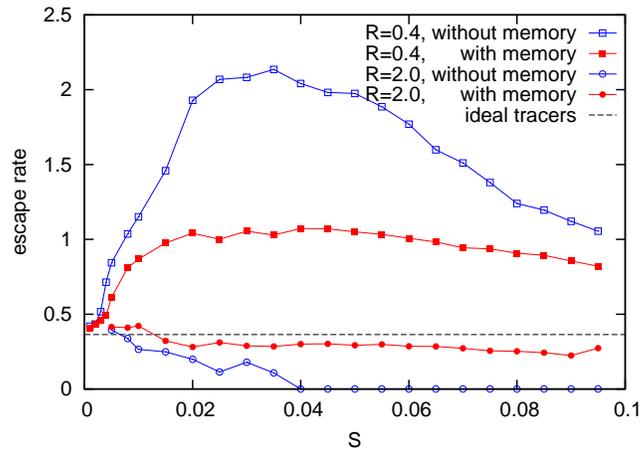}
\par\end{centering}

\caption{\label{fig:escaperates-cut}Escape rate as a function of $S$ with and without memory for
$R=0.4$ and $R=2$.}
\end{figure}

The escape rates are typically closer to that of the passive (neurally
buoyant) case with memory than without.  In other words, the speed of
the emptying process becomes less $R$-dependent.  Figure~\ref{fig:escaperates-cut} illustrates this tendency along two cuts (at
$R=0.4$ and $R=2$). It is clear that the escape rates of aerosols
(bubbles) become smaller (larger) due to memory, and that they are
typically above (below) that of the passive case. There might be,
however, exceptions as the bubble case with $S \le 0.01$ illustrates where
the escape rates slightly exceeds $\kappa_{\mm{ideal}}$. Note that the
$S$-dependence is also weaker with memory. The escape process depends, thus, less
on both parameters $R$ and $S$ when memory is taken into account.

\begin{figure}
\noindent \begin{centering}
\includegraphics[width=0.5\columnwidth]{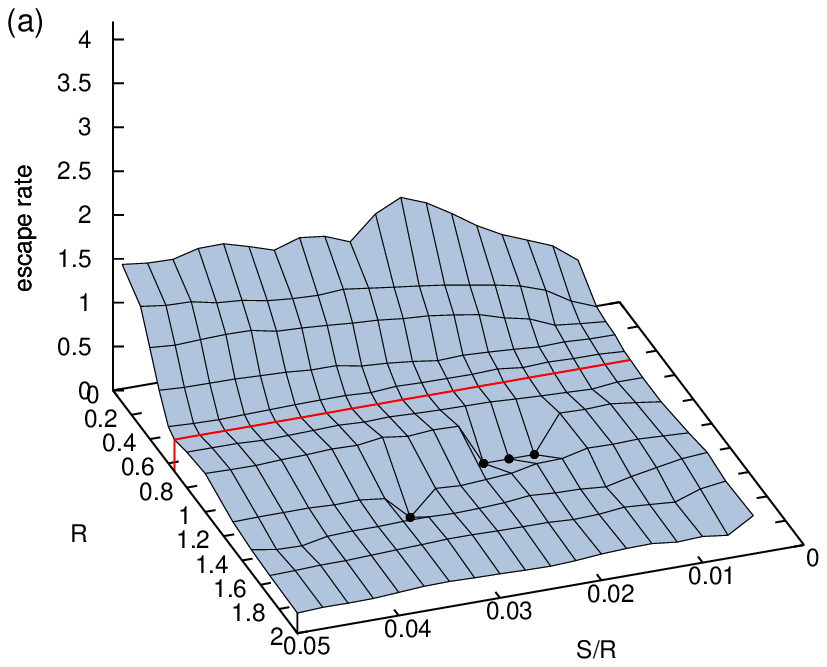}\includegraphics[width=0.5\columnwidth]{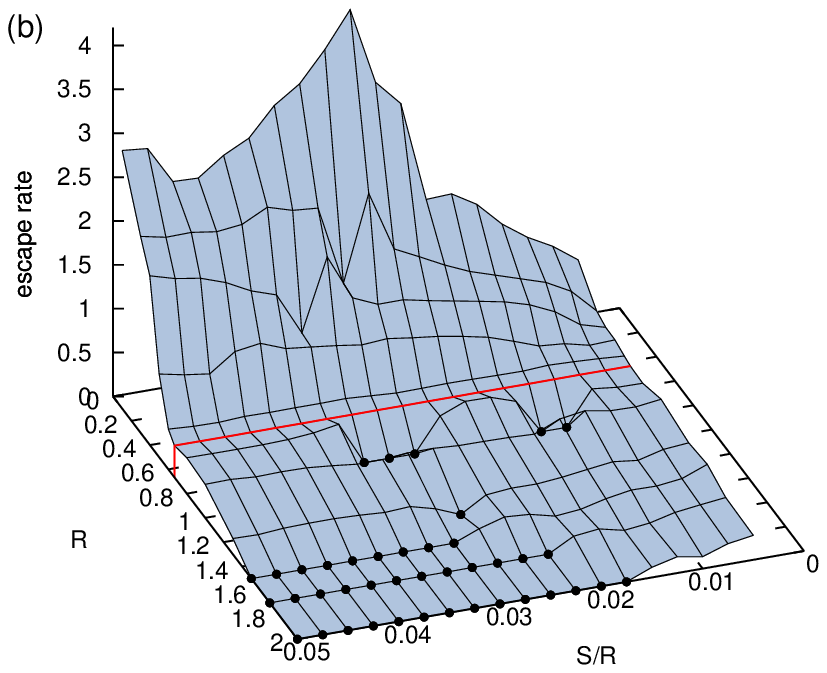}
\par\end{centering}
\caption{\label{fig:escaperates-A} The same as
  figure~\ref{fig:escaperates-S} on the parameter plane $(St=S/R,R)$.}
\end{figure}

For completeness, we also present the escape rates on the plane of the
Stokes number $St=S/R$ and the density ratio $R$,
figure~\ref{fig:escaperates-A}. Since in this representation the density
plays a role along both axes, the graph of the escape rate is
different but the general content remains the same.

\begin{figure}
\noindent \begin{centering}
\includegraphics[width=6.6cm]{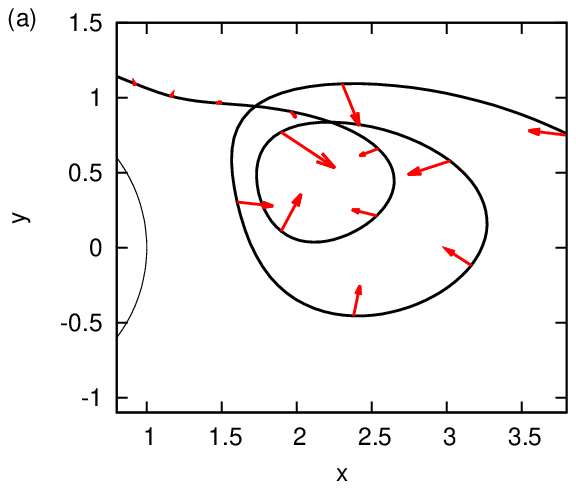}\includegraphics[width=6.6cm]{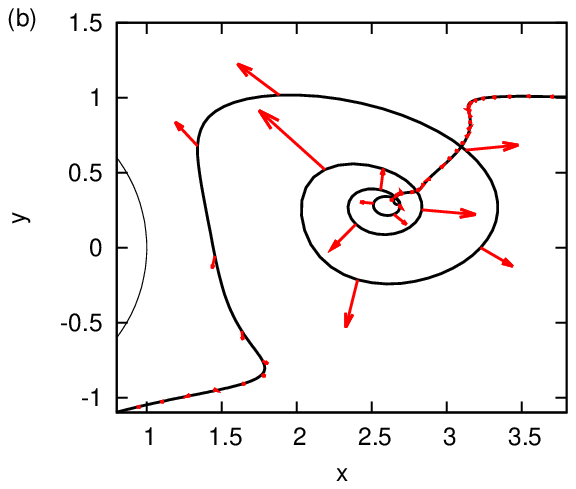}
\par\end{centering}

\caption{\label{fig:memory-force-in-vortex}A comparison of the history
  force (red arrows) acting on an aerosol (a) and a bubble (b) along a
  spiraling trajectory in the wake.  The parameters are $S=0.01$ and
  $t_{0}=0$ in both cases and $R=0.4$, $\vek
  r_{0}=\left(-0.5,1.2\right)$ for the aerosol and $R=2$, $\vek
  r_{0}=\left(-0.5,-1.14\right)$ for the bubble. The scale of the
  arrows for the bubble case is 10 time larger, i.e., in this case the
  history force is much stronger (this is also true for the the other
  forces). Note that the history force counteracts the centrifugal
  force in both cases.}
\end{figure}

An important effect of the history force can be observed when studying
particles moving along general curved trajectories: the history force
is found to {\em counteract the centrifugal force}. Heavy particles
are pushed outwards of vortex centers by the centrifugal force. Memory
is seen to generate a dissipative effect which provides a force
pointing towards the center of a vortex, as illustrated by figure~\ref{fig:memory-force-in-vortex}a.  
For bubbles the effect is
opposite, i.e., the history force points outwards the center of a
vortex; it counteracts the (anti)centrifugal force which pushes bubbles
inside of vortices. This behavior has been also found in
\cite{Angilella2004} for the analytically treatable case of a single
vortex with the flow field $\vek u(\vek r)=\left|\vek r\right|\vek
e_{\varphi}$. The effect visualized in figure \ref{fig:memory-force-in-vortex} provides a qualitative explanation for the
decrease (increase) of the escape rate for heavy (light) inertial
particles in the presence of memory.

We can summarize that, in general, the history force tends to weaken
the difference between particle and fluid and ``pushes'' the dynamics
towards that of the ideal case.

\section{Residence time and non-hyperbolic behavior\label{sec:residence-times}}

\begin{figure}
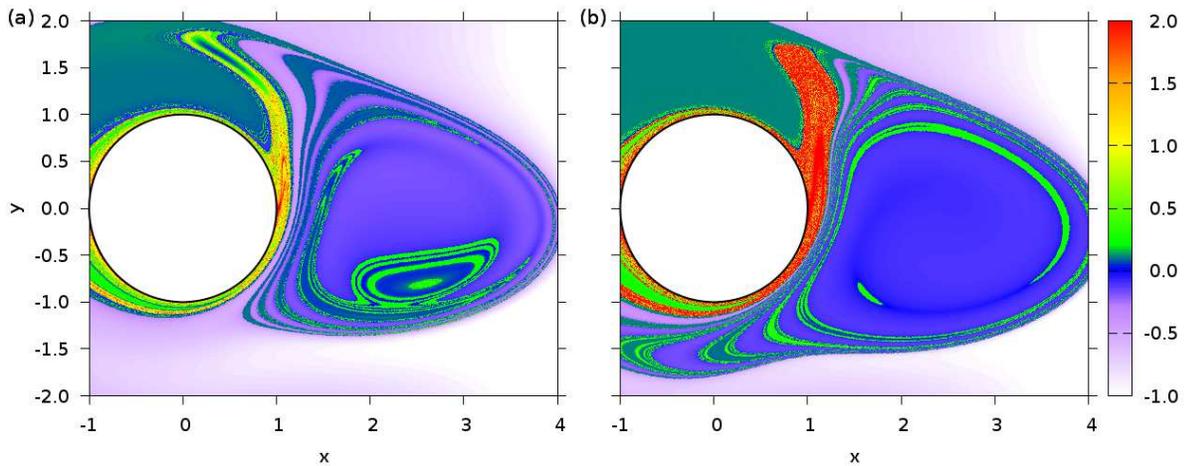

\begin{centering}
\includegraphics[height=0.4\columnwidth]{{{restime_R2.0_S0.02}}}
\par\end{centering}

\caption{\label{fig:residence-bubbles}Residence times with (a) and
  without (b) memory for $R=2.0$, $S=0.02$, $t_{0}=0$. The escape condition
  is $x>5$ ($r<1.014$ is not counted here as an escape). The logarithm ($\log_{10}$) of the
  residence time is color-coded.}
\end{figure}

A further characterization of the escape dynamics is via the residence
time distribution. To any initial position in the flow, we determine
the time the particle takes to escape and indicate this time via
color-coding. The trajectories are integrated up to a maximum of $100$
time units, thus the largest residence time found is $100$. In this
section we do not cut out the circle of radius 1.014 along the
cylinder surface (unlike in the previous sections),
and the escape condition is simply $x>5$,
in order to also see the
effect of the stickiness of the wall. Figure~\ref{fig:residence-bubbles}
shows the result with and without memory for bubbles. Very long lifetimes, longer
than $100$ time units, would indicate basins of attraction of an
attractor. In this case no attractors exist and lifetimes on the order
of 100 mark slowly moving trajectories which will spend a long time in
the boundary layer close to the cylinder surface. We see clearly that the
area of initial conditions with long lifetimes (redish colors) is much lower with
memory than without. Thus the surface of the cylinder seems to be less
sticky when memory is included. Furthermore the
structure of the residence time distribution shows a clear dependence
on the presence of memory also away from the cylinder surface, e.g. in
the region $\left[1.5,3.5\right]\times\left[-1.5,0\right]$ where a
vortical pattern is present with memory.

\begin{figure}
\noindent \begin{centering}
\includegraphics[width=6.6cm]{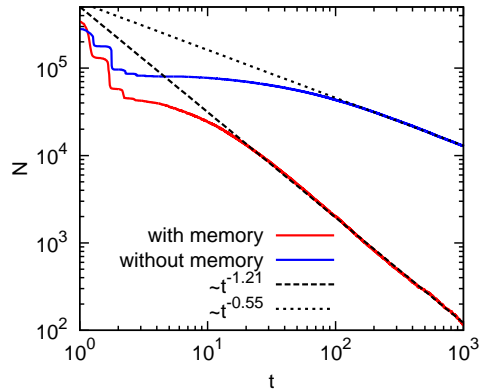}
\par\end{centering}

\caption{\label{fig:nonhyperbolic} Long time, non-hyperbolic regime of
  the escape dynamics for bubbles. Parameters are the same as in
  figure~\ref{fig:residence-bubbles}.}
\end{figure}

In dynamical system terms, the sticky surface of the cylinder leads to
a non-hyperbolic behavior marked by a long-term power law decay of the
number of survivors. For trajectory ensembles containing initial
points close to the surface one finds that the number N(t) of
non-escaped particles up to time t behaves for $t\gg1$ as
\begin{equation}
N(t)\sim t^{-\sigma},
\end{equation}
where $\sigma$ is the exponent of the non-hyperbolic decay
\cite{Lai2011}.  The number of survivors in the range of 100 and 1000
time units exhibits clear straight lines on a log-log plot in both
cases, as can be seen in figure~\ref{fig:nonhyperbolic}. The slopes are,
however, different. We have found a change from $\sigma=0.55$ to
$\sigma=1.21$ when memory is included. (For ideal tracers $\sigma=2$
\cite{Jung1993}.) For bubbles, memory effects lead thus to a faster
decay (to an increased decay exponent), weaken non-hyperbolicity, but
do not destroy the power-law behavior. For aerosols,
the effect of the cylinder wall appears to be
much weaker both with and without memory, and we don't find any change of the
non-hyperbolic decay exponent.

\section{Statistics of forces, slip velocity and acceleration\label{sec:PDFs}}

In section \ref{sec:Trajectories} we briefly discussed the forces
acting along a 
sample trajectory. To be able
to make more general statements we now turn to the statistics of these
forces. 
The statistics presented in this section are obtained by averaging
over an ensemble of $N_{0}\approx1.8\cdot10^{6}$ aerosols $(R=0.4, S=0.01)$,
started in the domain $\left[-1,4\right]\times\left[-2,2\right]$
at $t_{0}=0.2$; we sampled over all time instants at which a
particle has not yet escaped the wake of the cylinder.

\begin{figure}
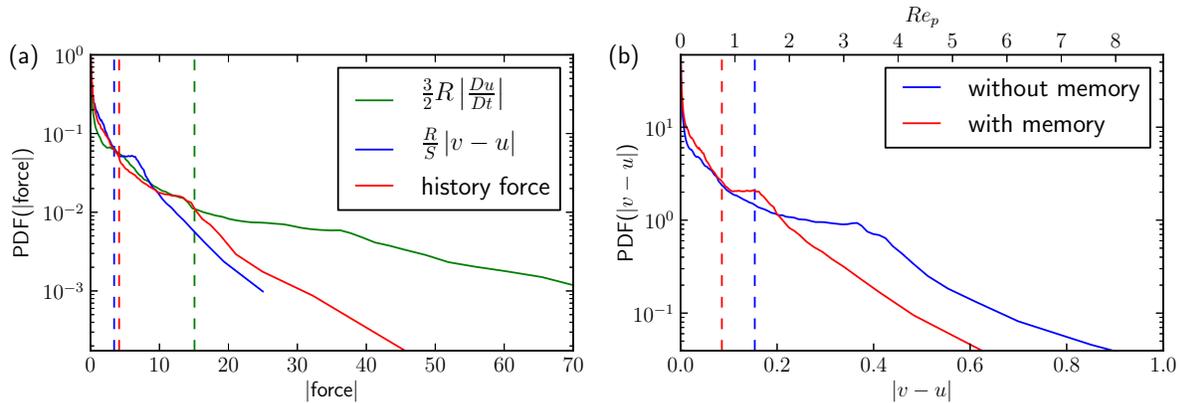

\noindent \begin{centering}
\includegraphics[width=0.5\columnwidth]{{{forcespdfs_R0.4_A40}}}\includegraphics[width=0.5\columnwidth]{{{velocitypdf_R0.4_A40}}} \par\end{centering}

\caption{\label{fig:pdfs}(a) The PDF of the magnitude of the forces on
  the rhs of (\ref{eq:MR-Equations-dimensionless}), with memory.\\ (b)
  PDF of the magnitude of the dimensionless slip velocity $\left|\vek
  v-\vek u\right|$ with and without memory. The x-axis on the top
  shows the particle Reynolds number $Re_{p}$, which is proportional
  to the slip velocity (see (\ref{eq:Re-slip-relation})).
  Vertical lines indicate the position of the corresponding
  averages.  The parameters are $R=0.4$, $S=0.01$. The
  ensemble consists of $N_{0}\approx1.8\cdot10^{6}$ particles initiated in the
  domain $\left[-1,4\right]\times\left[-2,2\right]$ around the
  cylinder at $t_{0}=0.2$; a particle is part of the ensemble
  only until it escapes.}
\end{figure}

Figure~\ref{fig:pdfs}a shows the probability density function (PDF) of
the magnitude of the three forces appearing on the right-hand side of (\ref{eq:MR-Equations-dimensionless}). Let us first compare the averages
of the different forces, shown as vertical lines in
figure~\ref{fig:pdfs}a.  We see that our observations from
section~\ref{sec:Trajectories} based on a single example trajectory also
hold true for the averages: the pressure term is the dominant one and the
history force (red curve in figure~\ref{fig:pdfs}a) is of similar magnitude as the Stokes drag.  Looking at
the tails of the PDFs we observe that the pressure term has a
particularly large amount of extreme events and the history force has a
slightly longer tail (more extreme events) then the Stokes drag.

The ratio of the averages of the history force and of the Stokes drag in
figure~\ref{fig:pdfs}a is $F_h/F_d=1.21$. In view of this finding, (\ref{eq:ratio}),
and the fact that $\sqrt{S}=0.1$, we conclude that parameter
$\alpha$ is $\sqrt{2/9}\times 12.1=5.7$ in this setting.  The observation that the $\alpha$ value is not close to
$1$ shows that the importance of the history force cannot be
estimated solely by $\sqrt{S}$; parameter $\alpha$ is also
relevant. If we neglected $\alpha$, we would underestimate the
history force by a considerable amount.

In order to be a useful parameter for estimating the importance of the
history force, $\alpha$ should be independent of $S$. Indeed, for
$R=0.4$ we find $\alpha\in[4.7,6.0]$ and for $R=2.0$
$\alpha\in[7.2,7.7]$ when $S$ is varied between $0.05$ and $0.005$. For
the purpose of rough estimates we can thus consider $\alpha$ constant,
with a value around $5$ (in this flow).

Figure~\ref{fig:pdfs}b shows the PDF of the magnitude of the slip
velocity $\left|\vek v-\vek u\right|$. It is important to see that the
average is much below $1$ in both cases, indicating that the deviation
between the particle and the flow velocity is typically much smaller
than the flow velocity itself.  Furthermore, the average dimensionless
slip velocity $0.153$ of the memoryless case is reduced to $0.085$ (to
ca. $55\%$) when the history force is included. Thus, particles with
memory follow the fluid {\em more closely}.  In harmony with this, the
tail of the slip velocity PDF becomes shorter with memory, i.e., there
are less strong ``detachments'' from the fluid flow. We find the same
tendency for bubbles albeit somewhat weaker, e.g. for $R=2$, $S=0.01$
the slip velocity is reduced to $85\%$ by memory.  All this
illustrates that the history force provides an important viscous
contribution so that the resultant of it and the drag ensures a faster
relaxation towards the flow velocity than the drag alone.

Equation~(\ref{eq:Re-slip-relation}) shows that the slip-velocity is
proportional to the instantaneous $Re_{p}$.  Accordingly, the top axis
of figure~\ref{fig:pdfs}b displays $Re_{p}$. Its average for particles
with (without) memory is found to be $0.76$ ($1.36$).  The assumption
that the particle Reynolds number remains of order unity or smaller is
thus found to be fulfilled on average, with a somewhat sharper
bound with memory.

The statistics of the acceleration of particles (not shown) is also
consistent with the picture above.  Ideal tracers have the highest
average acceleration and the most extreme events (the longest tail in
the PDF). The case without memory exhibits the lowest accelerations, and the
inclusion of memory increases the average acceleration, as well as, the
probability of extreme events. Thus we see again that the statistics
come closer to that of the ideal case when memory is included.

\section{Periodic attractors\label{sec:attractors}}

Let us now turn to a parameter set where attractors are present
both with and without memory: $w=24$ (instead of $192/\pi$),
where $w$ describes the strength of the vortices behind the cylinder,
and $R=1.5$, $S=0.03$. The residence times are depicted in
figure~\ref{fig:residence-bubble-w24}. Their structure changes
significantly when memory is taken into account, as we have already
seen in section \ref{sec:residence-times}. Here we included again the circle
of radius $1.014$ as an escape condition to exclude the non-hyperbolic
decay. Thus residence times of 100 (the maximum integration
time) show initial conditions leading to an attractor.  In the presence
of memory, the basin of attraction (marked by red) becomes
significantly smaller; its area is reduced by a factor of about $2.5$. Thus,
we can say that memory makes the attractor less attractive in this
case. We note, that for this choice of $w$ there are also trajectories
of ideal tracers which never leave the wake. Tiny integrable regions
exist which appear as red dots in
figure~\ref{fig:residence-bubble-w24}c. Comparing the residence time
plots we see again that the case with memory comes closer to that of
ideal tracers.

\begin{figure}
\begin{centering}
\includegraphics[height=0.4\columnwidth]{{{restime_R1.5_S0.03_w24}}}
\par\end{centering}

\caption{\label{fig:residence-bubble-w24}Residence times for bubbles
  with (a) and without memory (b), and for ideal tracers (c).
  The parameters are $R=1.5$, $S=0.03$, $t_{0}=0.3$, $w=24$. The
  logarithm ($\log_{10}$) of the residence time is color-coded.}
\end{figure}

\begin{figure}
\begin{centering}
\includegraphics[width=11cm]{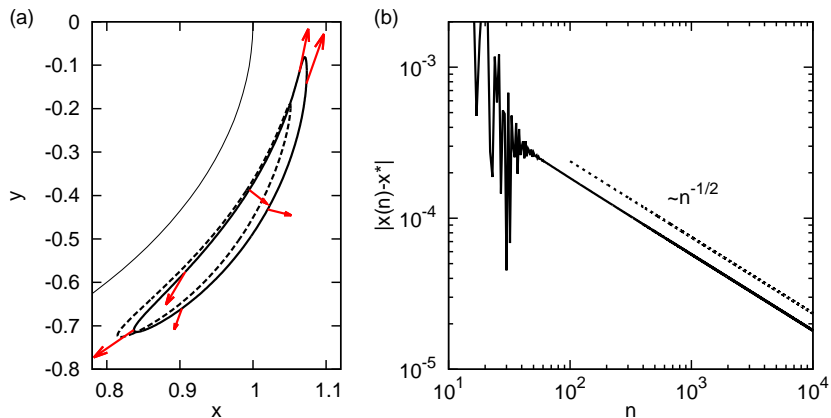}
\par\end{centering}

\caption{\label{fig:bubble-attractor-w24} (a) Attractors with and without
memory (solid and dashed line, respectively) traced out by approaching
trajectories after $10^{4}$ time units (the parameters are $R=1.5$,
$S=0.03$, $t_{0}=0.3$, $w=24$.). The arrows represent the history force.
(b) Distance between an approaching trajectory with memory and the
corresponding asymptotic attractor at integer times. }
\end{figure}

It is worth comparing the attractors with and without memory. Both
attracting objects appear to be periodic orbits of period-1 (i.e.,
with the same period as the vortex shedding period), and are of
similar form, as figure~\ref{fig:bubble-attractor-w24}a shows. Note that
the history force does not vanish on the attractor with memory.  A
basic difference between the two cases is that the attractor of the
memoryless case is stationary by the time the plot is made, but the
cycle-looking object seen in the presence of memory keeps changing
very slowly in time.  It is thus a 'quasi-attractor' only.  A real
attractor is expected to be reached (even in a numerical sense) after
a very long time only.

We examine the convergence of the trajectory to the attractor through
the quantity $\left|x(n)-x^{*}\right|$, where $x(n)$ and $x^{*}$ is
the $x$-coordinate of an approaching trajectory after $n$ time units and of the asymptotic
attractor ($x^{*}=\lim_{n \rightarrow \infty}x(n)$),
respectively. This corresponds to taking snapshots at integer
multiples $n$ of the period, i.e., a stroboscopic map. The time
dependence of this quantity is shown in
figure~\ref{fig:bubble-attractor-w24}b for the case with memory. The
convergence follows a $t^{-1/2}$ law, where the exponent is obviously
related to the power law behavior of the kernel of the history force
(\ref{eq:memory_derivative}).
The algebraic convergence is in strong contrast to normal dynamical
systems where the convergence is exponential. This means that
with memory the convergence is \emph{very} slow and that there is no
characteristic time scale characterizing this convergence.

The fact that a periodic attractor is possible with memory can be
explained by the observation that after a very long time the memory of
the approach to the attractor decays away, and the dynamics remembers
only what has been in the close vicinity of the attractor,
on a loop like the continuous line in figure~\ref{fig:bubble-attractor-w24}a, and a
convergence becomes thus possible.

\section{Interpretation in terms of snapshot attractors\label{sec:snapshot-attractors}}

Using this classical single trajectory picture, one concludes that
the attractor can be reached after a very long period of time only.
There is, however, an alternative view, that of particle ensembles,
also available. In autonomous systems the two views are equivalent,
which is not so obvious in non-autonomous problems, like ours (see (\ref{eq:strobm})).
In this class, one can then define a snapshot attractor as an object which attracts
all the trajectories initialized in the infinitely remote past within
a basin of attraction \cite{Romeiras1990,Ghil2008}. It can be obtained
by monitoring an ensemble of particle trajectories all subject to
the same non-autonomous equation of motion. After a characteristic
dissipative time, the ensemble traces out a snapshot attractor. This
attractor might, however, move continuously in time.

In the dynamical systems community, the concept has been known for
many years \cite{Romeiras1990}. The idea proved to be particularly
well suited for understanding the advection of passive particles in
random flows \cite{Jacobs1997,Neufeld1998,Hansen1998}, and explains
experimental findings \cite{Sommerer1993}. More recently, snapshot
attractors have turned out to be promising tools to describe the variability
of climate dynamics in a novel way \cite{Ghil2008,Checkroun2011,Bodai2011,Bodai2013,Ghil2013}.
In these settings the non-autonomous driving is typically a random
noise or a sustained chaotic process. The snapshot attractor is then
ever changing, and appears to be a fractal object whose shape is evolving
in time. In cases with an eventually vanishing driving, the snapshot attractor
might have a time-dependence that ceases asymptotically.

\begin{figure}
\begin{centering}
\includegraphics[width=0.9\columnwidth]{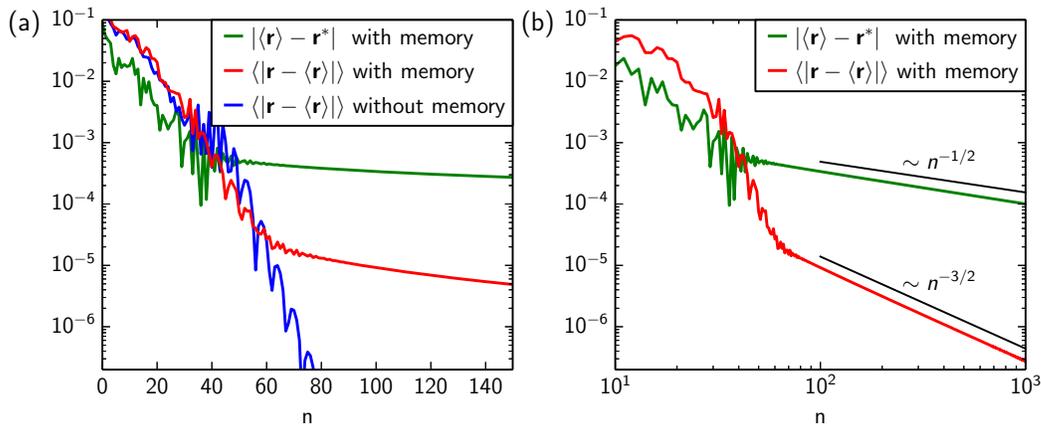}
\par\end{centering}

\caption{\label{fig:ensemble-convergence} Convergence of an ensembles
  of 100 particles towards a fixed point attractor on the stroboscopic
  map with memory and without memory. The ensemble consists of 100
  randomly chosen initial conditions from the basin of attraction in
  the region $[0.5,1.5]\times[-1,1]$. The green curve represents the
  distance of the center of mass from the true, asymptotic attractor
  $\vec r^*$ (an analog of figure~\ref{fig:bubble-attractor-w24}b).  The
  other curves represent the average radius of the ensemble (the red
  one with, the blue one without memory).  Panels (a) and (b) are
  log-lin and log-log representations and focus on the short and long
  time approach, respectively. The parameters are $R=1.5$, $S=0.03$,
  $t_{0}=0.3$, $w=24$.}
\end{figure}

For our problem of chaotic advection with memory, where a slow
convergence towards a traditional periodic attractor takes place, one can
assume that close to this attractor the effect of the history force
for neighboring members of the particle ensemble can locally be considered as a
non-autonomous
perturbation superimposed on the usual inertial particle dynamics. An
ensemble of particles might then converge to a snapshot attractor, a
fixed point on the stroboscopic map, after some time. This snapshot
fixed point attractor is then slowly shifted towards the asymptotic
traditional attractor. Here a separation of time scales is expected to
occur since the convergence to the snapshot attractor is a usual
dissipative effect, and hence this decay should be exponential,
whereas the slow shift is due mainly to the diffusive kernel, and follows
thus a power law.

To verify this we choose an ensemble of $100$ particles approaching
the periodic attractor, and compute the center of mass $\left\langle
\v r(n)\right\rangle$ (i.e., the average of the spatial location
vector over the ensemble) and the radius of the ensemble $\left\langle
\left|\v r(n)-\left\langle \v r(n)\right\rangle \right|\right\rangle$
at integer times $n$. The center of mass represents the time-dependent
position of the snapshot attractor whereas the radius is a measure of
convergence to the snapshot attractor. The center of mass is found to
exhibit the same power-law behavior, of power $-1/2$, as
$x(n)$. However, we observe that the radius of the ensemble has an
initial roughly exponential decay, see
figure~\ref{fig:ensemble-convergence}a, which crosses over to an
algebraic decay of power $-3/2$. Thus we conclude that the point-like
object formed after about 70 time units is a snapshot fixed point
attractor emerging as an effect of memory.

\begin{figure}
\begin{centering}
\includegraphics[width=0.715\columnwidth]{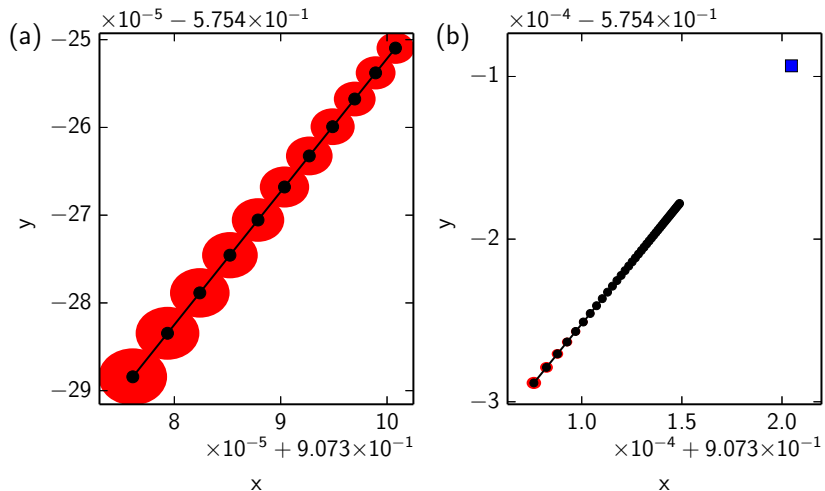}
\par\end{centering}

\caption{\label{fig:ensemble-jumps} Positions of an ensemble of 100
  particles (the same as in figure~\ref{fig:ensemble-convergence})
  approaching a fixed point attractor on the stroboscopic map.  Black
  dots and red circles around them represent the center of mass and
  the radius, respectively, of the ensemble. The parameters are the
  same as in figure~\ref{fig:ensemble-convergence}. (a) The positions
  of the ensemble at time instants $200$, $210$, ..., $300$. (b) The
  positions of the ensemble at time instants $200$, $220$, ...,
  $1000$. The circles indicating the radius are hardly visible at this
  scale. The asymptotic fixed point attractor is marked by a
  square. Note the largely magnified scales in both panels.}
\end{figure}

To deepen this picture, we plot in figure~\ref{fig:ensemble-jumps} the
center of mass and the radius of the ensemble in the plane of the
fluid at integer times.  Figure~\ref{fig:ensemble-jumps}a shows time
instants between $n=200$ and $300$, whereas
figure~\ref{fig:ensemble-jumps}b covers the range $[200,1000]$. The
scale is strongly magnified but the plot clearly indicates that the
fixpoint snapshot attractor slowly moves.  Even after $1000$ time
units it is still away from the limiting location.  The latter is
determined from a self-consistent fit to the algebraic decay process
of the center of mass.

The evolution can thus be split into a short-time and a long-time
regime. In the first one, up to about $70$ time units, a convergence
to a snapshot attractor takes place. In the second one (for $n>70$) a
snapshot attractor is reached but it is shifted slowly towards an
asymptotic fixed point, the traditional attractor.

\section{Chaotic attractors}\label{sec:chaoticattr}

\begin{figure}
\begin{centering}
\includegraphics[height=0.4\columnwidth]{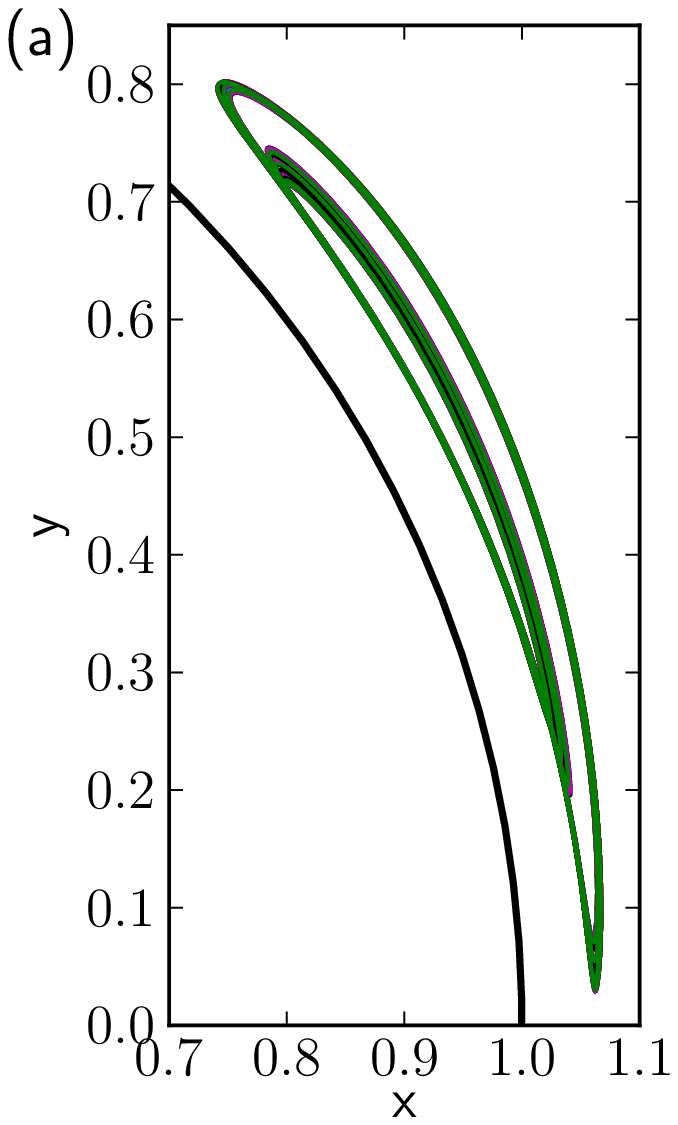}
\includegraphics[height=0.4\columnwidth]{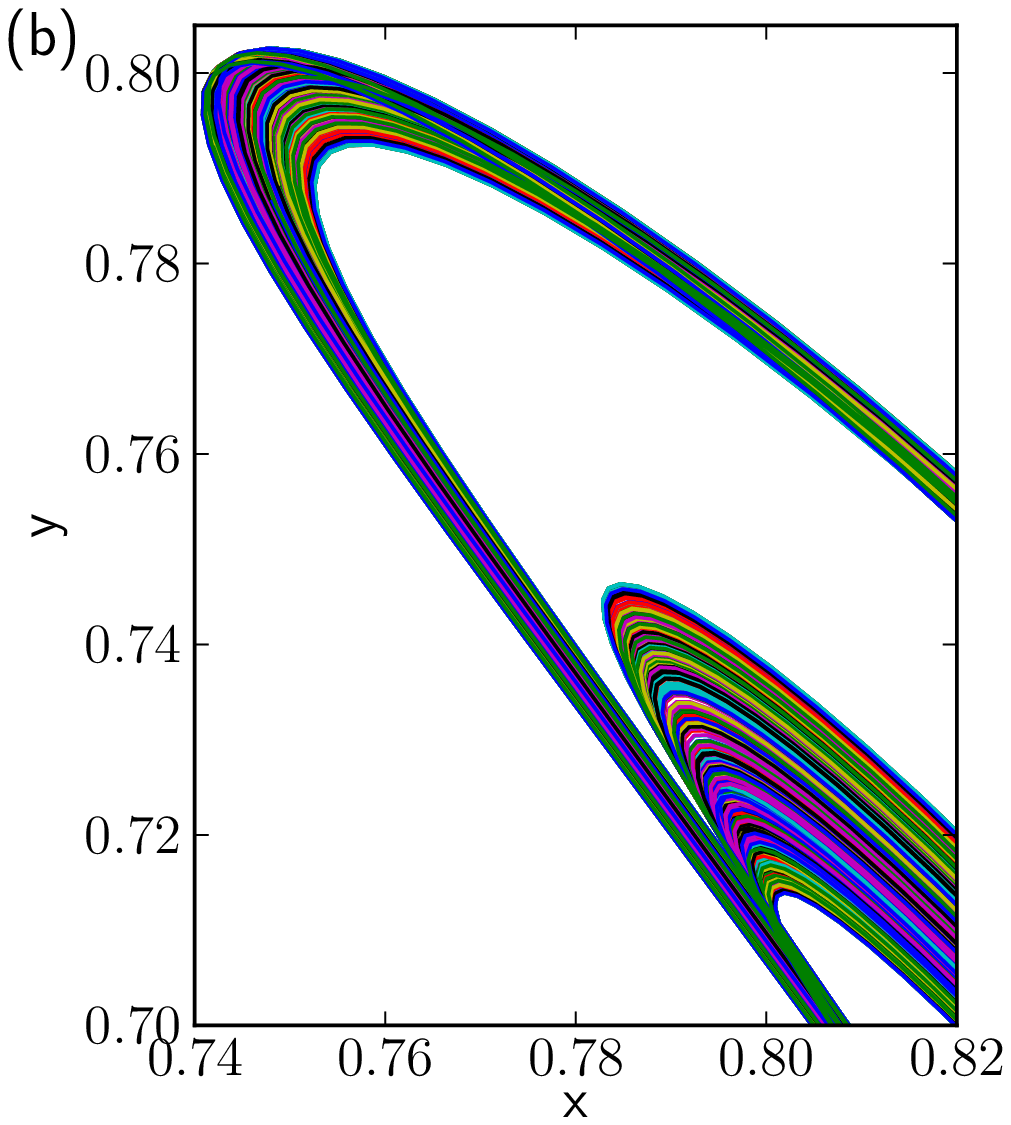}
\includegraphics[height=0.4\columnwidth]{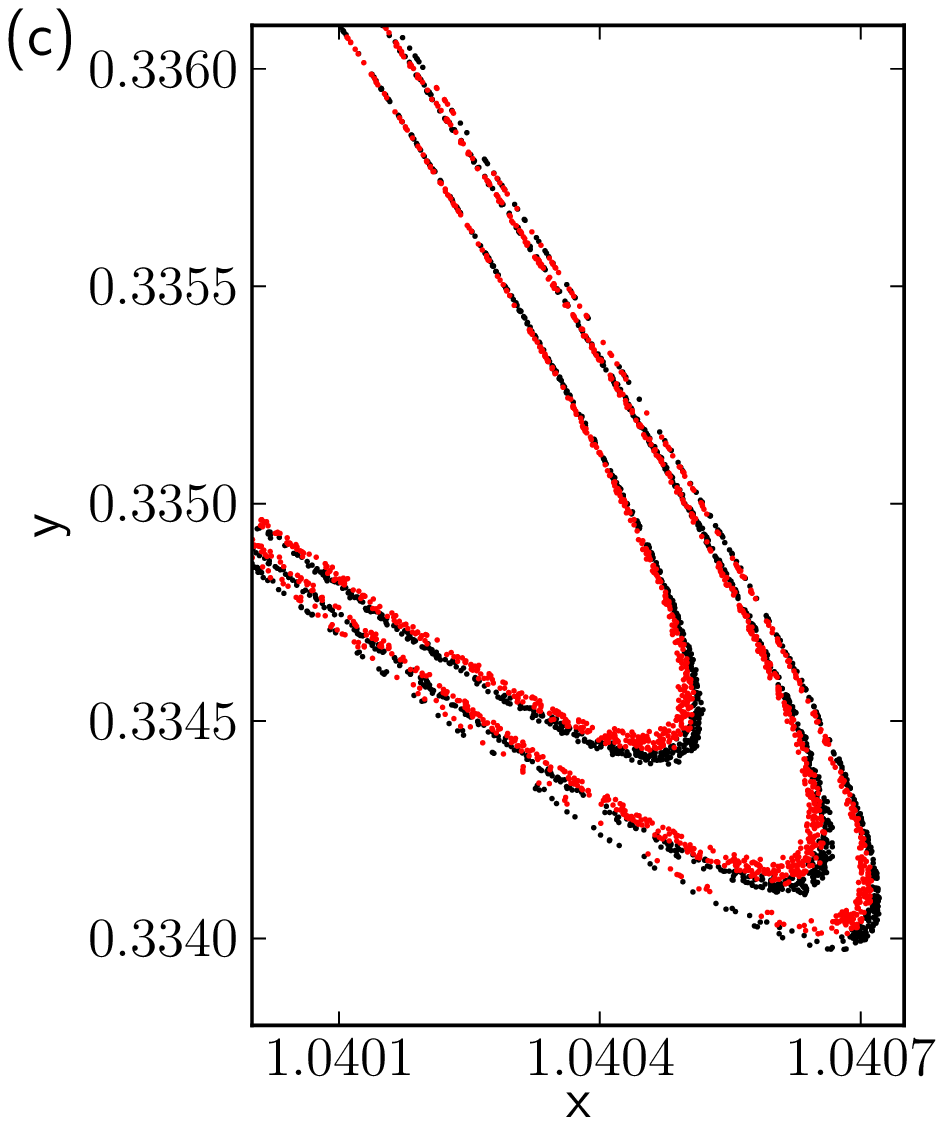}
\par\end{centering}

\caption{\label{fig:chaotic-attractor}Chaotic attractor with memory.
  (a) Positions of an ensemble of 100 trajectories (marked by
  different colors) traced out in the time interval $t\in[990,1000]$;
  (b) A magnification.  The ensemble consists of 100 randomly chosen
  initial conditions from the basin of attraction in the region
  $[0.5,1.5]\times[-1,1]$. The parameters are $R=1.47$, $S=0.049$,
  $t_{0}=0.3$, $w=31.2$.  (c) Positions on a stroboscopic map (on a
  magnified scale). The red dots correspond to the attractor points at
  $n\in[600,800]$ and the black dots to those at $n\in[800,1000]$.  We
  observe a slow motion of the attractor obtained in the ensemble
  picture.  }
\end{figure}

Chaotic attractors with memory are very rare in the parameter range
investigated and we had to perform an extended parameter-scan to find
one. In figure~\ref{fig:chaotic-attractor}a we see an ensemble of 100
trajectories (in continuous time) tracing out this chaotic attractor
with memory. A magnification, figure~\ref{fig:chaotic-attractor}b,
indicates clear fractal properties. The Lyapunov exponent of this
attractor is $\lambda=0.18\pm0.01$. It is positive, which shows that
the attractor is chaotic indeed.  Details of how the largest Lyapunov
exponent can be determined in such a system with long memory is given
in \ref{sec:appendix-lyapunov}. We note that at this parameter values
(given in the caption to figure~\ref{fig:chaotic-attractor}) the attractor without
memory is not chaotic.

Figure~\ref{fig:chaotic-attractor}c exhibits the stroboscopic picture of
the chaotic attractor with memory. It shows a magnification in two
different time intervals: $t\in[600,800]$ and $t\in[800,1000]$.  A
shift on the order of $10^{-4}$ units can be observed, a clear
manifestation of the fact that this chaotic attractor is slowly
drifting, similar to the fixed point snapshot attractor discussed in
the previous section.

\section{Chaotic saddles}\label{sec:saddle}

In contrast to permanent chaos,
transient chaos is ubiquitous in the problem as the typical form of
chaos in this open flow.
Chaotic transients are found at any parameter,
irrespective of the existence of attractors. The degree of chaos can
thus best be compared by comparing the underlying chaotic saddles which
exist both for ideal tracers and inertial particles with or without memory.

\begin{figure}
\noindent \begin{centering}
\includegraphics[width=\columnwidth]{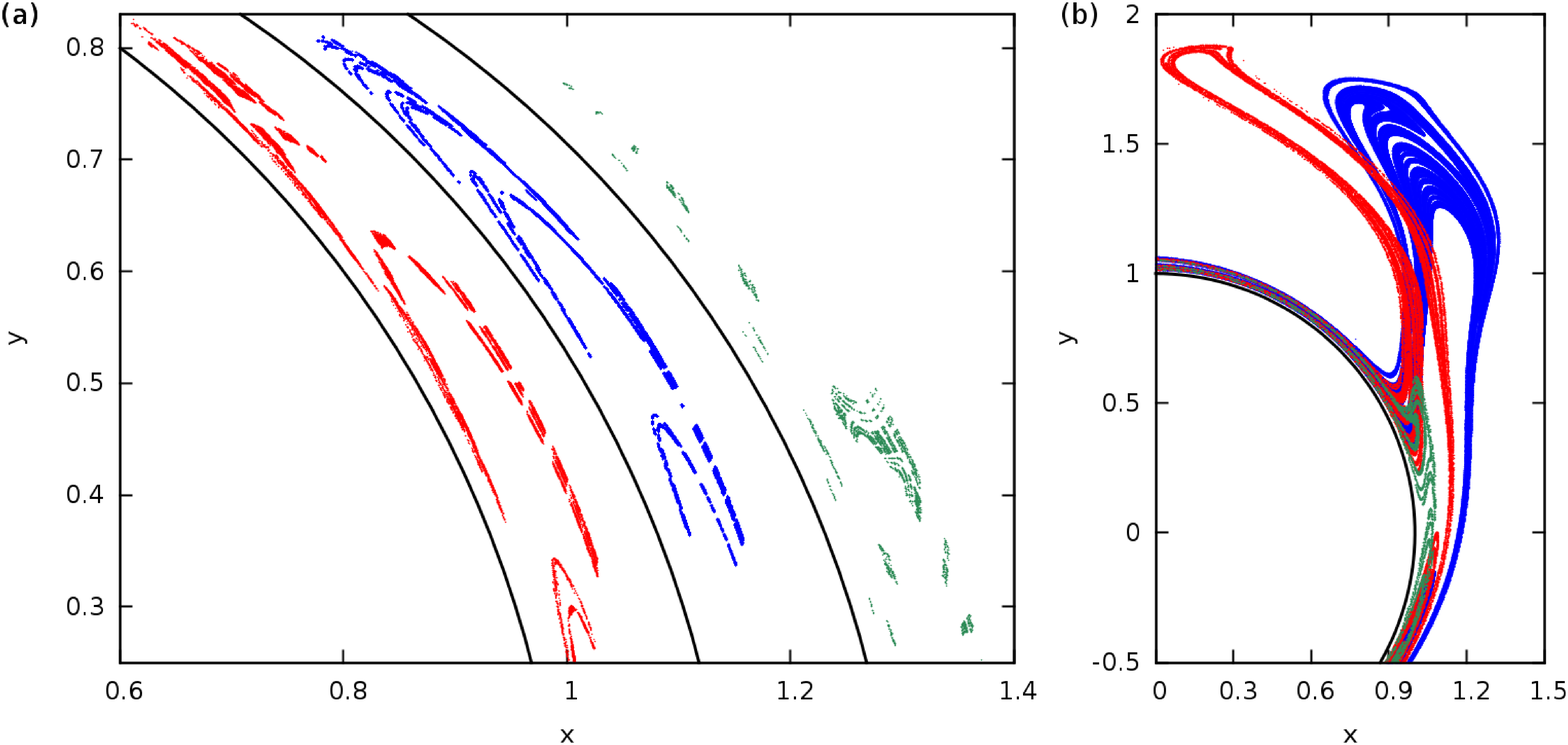}
\par\end{centering}

\caption{(a) The chaotic saddle for inertial particles with memory (red), without memory (blue), and for ideal tracers (green). The saddle
without memory and the one for ideal tracers are shifted horizontally by 0.1 and 0.2, respectively, for better visibility.
(b) The stable manifolds of the corresponding saddles.
The parameters are $R=2.0$, $S=0.02$, $t_{0}=0$, the same as in figure~\ref{fig:residence-bubbles}.
}
\label{fig:saddle}
\end{figure}

To construct the saddle, we consider particles with long residence
times, longer than $15$ time units. The idea is that these particles
come close to the saddle and then leave the wake. The longer the
particle stays in the wake, the closer it comes to the saddle. One
might suppose that at half of their residence time ($t=7$) the
particles are closest to the saddle and thus their positions at this
time trace out the saddle. This is supported by the observation that this
set of points at $t=7$ is only marginally different from those at
$t=6$ and $t=8$, i.e., the set is (approximately)
invariant. Figure~\ref{fig:saddle}a shows the saddle constructed by this
method for three different advective dynamics.  The initial positions
of particles which stay long in the flow and thus come close to
the saddle trace out the stable manifold of the chaotic saddle, and are plotted in
figure~\ref{fig:saddle}b.

We see that the saddle and its stable manifold change considerably
when the history force is taken into account. It is interesting to
note that the saddle and its stable manifold differ much more from those of
ideal tracers than from the memoryless case. This illustrates that the general
rule according to which the presence of memory makes the dynamics to
be similar to that of ideal tracers is not true in all possible
aspects.

To further characterize the saddle and the corresponding dynamics we
calculated the Lyapunov exponent,
as
described in \ref{sec:appendix-lyapunov}. The Lyapunov
exponent with the history force is found to be $\lambda=0.91\pm
0.02$. This is larger then without memory, when $\lambda=0.79\pm
0.02$, and is close to that of ideal tracers for which
$\lambda=0.92\pm 0.02$. Thus particles with memory are
dynamically more unstable, and their measure of instability is closer to that of ideal tracers.

\section{Summary and discussion} \label{sec:discussion}

In summary, memory effects can have a very significant influence on inertial
particles. In the presence of the history force,
compared to the memoryless inertial case, we find:
\begin{itemize}
\item
a rare appearance of attractors, especially of chaotic ones,
\item
a less frequent appearance of caustics,
\item
a decrease of the centrifugal effect,
\item
a slower (faster) escape for aerosols (bubbles),
\item
a weaker effect of the cylinder wall,
\item
a stronger dynamical instability on chaotic sets,
\item
a very slow, algebraic convergence to attractors (when defined in the usual single trajectory picture),
\item
a tendency of the particles to follow the flow more closely.
\end{itemize}
As a trend, the dynamics become more similar to that of ideal tracers
when the history force is included (although exceptions can also be
found, see, e.g., figure~\ref{fig:saddle}).

In one principal aspect the dynamics basically differs, however, from
both the usual inertial and the passive tracer dynamics. Due to
memory, the problem is strictly speaking high-dimensional,
and methods known from low-dimensional dynamical systems are not
necessarily applicable.  For understanding the slow convergence
towards periodic attractors, the concept of snapshot attractors
appears to be useful.  More generally, we assume that the snapshot
interpretation, i.e., considering the instantaneous position of an
ensemble of trajectories started in the remote past, is applicable for
any kind of asymptotic sets (chaotic attractors, chaotic saddles) in
advection with memory effects. In this new picture, an exponential
convergence to a snapshot-like attractor is expected to show up,
which is then slowly changing afterwards.

Let us come back to estimation (\ref{eq:ratio}) of the importance of
the history force. A question of relevance is whether one can
determine a typical value for $\alpha$ in general. For the following
argument we assume the time to be dimensional again.  Imagine a signal
$f(t)$ which varies on the time scale $\tau$. Then we can estimate the
magnitude of the derivative $\dt f$ by $|f|/\tau$.  Extrapolating this
reasoning to fractional derivatives, we expect the magnitude of
$\left(\dt{}\right)^{1/2}f$ to be roughly $\sqrt{1/\tau}|f|$.  Thus,
setting $f=\vek v-\vek u$ 
in (\ref{eq:beta})
we obtain%
\footnote{The factor $\sqrt{T}$ appears because time is dimensional
  here in contrast to (\ref{eq:beta}).
}
\begin{equation}
\alpha\equiv\frac{\sqrt{T}\left(\dt{}\right)^{1/2}(\vek v-\vek u)}{\vek v-\vek u}\approx\sqrt{\frac{T}{\tau}}.
\label{eq:betaest}
\end{equation}
In our flow the time unit $T$ is chosen, traditionally, as the shedding period of the vortices. As particles
``live'' on the small scales, $\tau$ should be the smallest time scale
of the flow%
\footnote{In principle the typical time scale of $\vek v - \vek u$
  depends on the particle parameters. Nevertheless, it should be close
  to the small time scale of the flow. We choose this time scale for
  our order of magnitude estimates. This choice is further supported
  by the following finding: using (\ref{eq:expansion}) one can obtain
  a Taylor-expansion of $\alpha$ (\ref{eq:beta}) in $S$, where the
  zeroth order approximation
  $\alpha\approx\left|\frac{\mm{D}^{1/2}}{\mm{D}t^{1/2}}\Dt{\vek
    u}\right|/\left|\Dt{\vek u}\right|$ is solely a property of the
  flow (here $(\mm{D}/\mm{D}t)^{1/2}$ is defined in the same way as
  $(\mm{d}/\mm{d}t)^{1/2}$ but with the derivative taken along a fluid
  element trajectory).}.  In the von K\'arm\'an problem there are two
characteristic times: the shedding period $T$ and the convective time
scale $L/U$. The later is the smaller one and can be considered to be
the time a particle needs for going around a vortex. We thus take
$\tau=L/U$ in (\ref{eq:betaest}) and find, in view of (\ref{eq:Sl}),
that
\begin{equation}
\alpha\approx\sqrt{\frac{TU}{L}}=Sl^{-1/2}.
\end{equation}
Using the value $Sl=1/14$ we obtain $\alpha\approx3.7$, which is
quite close to the measured range $4.7\leq\alpha\leq7.7$.

The reason $\alpha$ turns out to depend on $Sl$ is that we have used the shedding period $T$ to
nondimensionalize the equation of motion, rather than the smaller time scale
$\tau=L/U$. Thus we conclude, a posteriori, that a more natural choice is to use the convective
time scale from the very beginning. Denoting the corresponding quantities in this time unit with
a prime, we obtain $\alpha'\approx1$ and

\begin{equation}
\frac{F_{h}}{F_{d}}\approx\sqrt{\frac{9}{2}S'}
\label{eq:ratio'}
\end{equation}
with a modified size parameter
\begin{equation}
S'=\frac{2}{9}\frac{a^{2}/\nu}{\tau}.
\label{eq:S'}
\end{equation}
We thus see that for the size parameter (and the Stokes number) --- a
Lagrangian quantity --- the use of the {\em smallest} time scale, the
convective time scale $L/U$, is natural.  It leads to simpler
expressions, in contrast to using the shedding period $T$ which is the
usual choice in the Eulerian characterization of this flow.

Let us come to a more general statement about other flows.  From
(\ref{eq:ratio'}) and (\ref{eq:S'}) we obtain the rather simple
formula
\begin{equation}
\frac{F_{h}}{F_{d}}\approx\frac{a}{\sqrt{\tau\nu}}
\label{eq:ratio-general}
\end{equation}
for the estimation of the history force relative to the Stokes drag.
We expect this to be a good estimation also in other flows as long as the time scale
$\tau$ is chosen to be the smallest time scale of the flow. This
choice becomes particularly important in multi-scale flows. Note also
that the convective time scale need not necessarily be the
smallest time scale in other flows. However, when this is fulfilled
and when the flow is smooth with only one characteristic
length scale $L$ (like the flow examined in this paper) we get
\begin{equation}
\frac{F_{h}}{F_{d}}\approx \frac{a}{L}\sqrt{Re}.
\end{equation}

Let us finally remark on the possible role of memory
effects in turbulent flows. In turbulence, the time scale $\tau$ should
be the Kolmogorov time scale $\tau_{K}$ \cite{Pope2000}. Using $\nu\tau_{K}=\eta^{2}$ in (\ref{eq:ratio-general}),
where $\eta$ is the Kolmogorov length, we obtain
\begin{equation}
\frac{F_{h}}{F_{d}}\approx\frac{a}{\eta}
\end{equation}
as an estimate for the importance of the history force.  We thus
expect memory effects to be negligible for $a<10^{-2}\eta$.  For $a$
around $0.1\eta$ the history force should be around $10\%$ of the
Stokes drag and is thus expected to be clearly observable.  However,
when $a$ is close to $\eta$, the history force can be as important as
the Stokes drag indicating strong memory effects even in turbulent
flows\footnote{For air turbulence in clouds $\eta$ is around
  $0.8\,\mathrm{mm}$ \cite{Grabowski2013} and for wind-driven oceanic
  turbulence it is in the range $0.3\,\mathrm{mm}-2\,\mathrm{mm}$
  \cite{Jimenez1997}.}. (Note that in the latter case Faxén corrections might also be
     important.)

\section{Acknowledgments}

Useful discussions with J.-R.~Angilella, T. Bohr, I.~Benczik, B.
Eckhardt, M. Farazmand, U.~ Feudel, K. Guseva, G. Haller, E. Ott,
A. Pumir and M. Wilczek are acknowledged. The project is supported by
the grant OTKA NK100296. We acknowledge support by the Alexander von
Humboldt Foundation, the Studienstiftung des deutschen Volkes, the
Deutsche Forschungsgemeinschaft, and the Open Access Publication Fund
of University of Muenster.

\appendix
\section{Lyapunov exponents in the presence of the history force} \label{sec:appendix-lyapunov}

Lyapunov exponents characterize the evolution of infinitesimal perturbations
in time. For the full Maxey-Riley equation~(\ref{eq:MR-Equations-dimensionless}) the equations governing
this evolution are

\begin{eqnarray}
\frac{1}{R}\dt{}\delta\vek v & = & \frac{3}{2}\delta\vek r\cdot\nabla\Dt{\vek u}-\frac{1}{S}\delta\vek v-\frac{1}{S}\delta\vek r\cdot\nabla\vek u\label{eq:perturbuation-evolution}\\
 &  & -\sqrt{\frac{9}{2\pi}\frac{1}{S}}\dt{}\int_{t_{0}}^{t} \frac{1}{\sqrt{t-\tau}}
 \left(\delta\vek v-\delta\vek r\cdot\nabla\vek u\right)\,\mm d\tau,\nonumber \\
\dt{}\delta\vek r & = & \delta\vek v.\nonumber
\end{eqnarray}
Here $\delta\vek q(t)\equiv(\delta\vek r(t),\delta\vek v(t))$ is an
infinitesimal perturbation along a given trajectory $\vek
q(t)\equiv\left(\vek r(t),\vek v(t)\right)$.  Equation
(\ref{eq:perturbuation-evolution}) is an evolution equation with
memory. 
The (largest) Lyapunov exponent is then defined as
\begin{equation}
\lambda=\lim_{t\rightarrow\infty}\frac{1}{t-t_0}\ln\frac{\left|\delta\vek q(t)\right|}{\left|\delta\vek q(t_{0})\right|}.\label{eq:definition-lambda}
\end{equation}
We assume that this quantity is the same for (almost) all trajectories of
a chaotic set, as in standard dynamical systems theory  \cite{Ott2002}.

One way to compute $\lambda$ is to numerically integrate
(\ref{eq:perturbuation-evolution}).  However, often it is more
convenient to obtain the evolution of perturbations directly from
pairs of particles starting close to each other. To a given trajectory
$\vek q(t)$ one computes a perturbed trajectory $\vek q'(t)$ starting
from a perturbed initial condition
\begin{equation}
\vek q'(t_{0})=\vek q(t_{0})+\Delta\vek q_{0},
\end{equation}
where $\Delta\vek q_{0}$ is a small but finite perturbation, and
obtains the Lyapunov exponent from
\begin{equation}
\lambda=\lim_{t\rightarrow\infty}\frac{1}{t-t_0}\ln\frac{\left|\Delta\vek q(t)\right|}{\left|\Delta\vek q_{0}\right|}
\end{equation}
with
\begin{equation}
\Delta\vek q(t)=\vek q'(t)-\vek q(t).
\end{equation}
 For this to be close to the original definition of $\lambda$
 (\ref{eq:definition-lambda}) one has to make sure that $\Delta\vek
 q(t)$ is small for all $t$, i.e.,  $\Delta\vek q(t)$ evolves according
 to a linear evolution equation.  A naive method would be to choose
 $\Delta\vek q_{0}$ so small that for a given integration time
 $t_{\mm{end}}$ one can be sure that $\Delta\vek q(t)$ remains
 small. However $\Delta\vek q_{0}$ can not be too small due finite
 numerical precision, because then $\vek q'$ would be numerically
 indistinguishable from $\vek q$.

The approach we use is to start with a small but not too small
perturbation ($\left|\Delta\vek q_{0}\right|=\Delta q_{min}$) and
adjust $\vek q'$ by rescaling $\Delta\vek q$ when $\left|\Delta\vek
q(t)\right|$ becomes larger than a given threshold $\Delta q_{max}$.
This way we make sure that $\Delta\vek q$ remains small. The
procedure is as follows:
\begin{enumerate}
\item integrate $\vek q'$ up to a time $\tilde{t}$ when $\left|\Delta\vek q\right|=\Delta q_{max}$,
\item set $\vek q'(t):=\vek q(t)+\Delta\vek q(t)\frac{\Delta q_{min}}{\Delta q_{max}}$ for $t_0\leq t\leq\tilde{t}$, and proceed with the integration for $\tilde{t}<t$,
\item repeat step 1 and 2 until the desired integration time $t_{\mm{end}}$ is reached.
\end{enumerate}
This rescaling makes sense because $\vek q'$ and $\vek q$ are rather
close and thus $\Delta\vek q$ evolves according to the \emph{linear}
equation~(\ref{eq:perturbuation-evolution}). Therefore, at any
iteration of the above procedure, $\Delta\vek q(t)\Delta q_{min}/\Delta q_{max}$ is an (approximate)
solution of (\ref{eq:perturbuation-evolution}) and $\vek q'(t):=\vek q(t)+\Delta\vek q(t)\Delta q_{min}/\Delta q_{max}$ is a valid perturbed trajectory.

The key idea 
is that when the upper
bound $\Delta q_{max}$ is reached, we just choose another solution $\vek q'$
which starts at a smaller initial perturbation $\Delta\vek q_0$; the
new perturbation is smaller but proportional to the previous one. As
$\Delta\vek q$ evolves according to a linear equation, we can efficiently obtain the new
solution by simple rescaling.  Because we rescale the previous
solution and continue with the integration only for $\tilde{t}<t$, we
avoid computations with too small $\Delta\vek q$.

With the history force it is important to adjust $\vek q'$ for the
whole past $t_0\leq t\leq\tilde{t}$ because the evolution depends on
the whole past. Without memory one would only need to adjust $\vek
q(\tilde{t})$, which leads to a standard method of computing the
Lyapunov exponent \cite{Benettin1980, Ott2002}.

When a perturbation trajectory $\Delta\vek q$ has been computed up to $t=t_{\mm{end}}$, we
obtain the Lyapunov exponent from
\begin{equation}
\lambda\approx\frac{1}{t_{\mm{end}}-t_0}\ln\frac{\left|\Delta\vek q(t_{\mm{end}})\right|}{\left|\Delta\vek q_0\right|},
\label{eq:lyapunov-1}
\end{equation}
where $t_{\mm{end}}$ is a large integration time.

In section \ref{sec:saddle} we are dealing with transient
chaos. Therefore the length of the trajectories is limited and it is
difficult to accurately compute the Lyapunov exponent from a single
trajectory. Therefore we use a preselected ensemble of trajectories which stay close to the saddle for a sufficiently long time, as in other cases of transient chaos \cite{Lai2011}. Equation~(\ref{eq:lyapunov-1}) 
defines the Lyapunov exponent as the slope of the function $\ln \left|\Delta\vek q(t)\right|$, computed using its values at $t=t_0$
and $t=t_{\mm{end}}$. We extend this idea to an ensemble by computing the
average
\begin{equation}
d(t)=\left\langle\ln\frac{\left|\Delta\vek q(t)\right|}{\left|\Delta\vek q_0\right|}\right\rangle
\end{equation}
and making a linear fit to $d(t)$ to obtain the slope which is the
Lyapunov exponent $\lambda$. In section \ref{sec:saddle} we choose
trajectories which stay in the wake for at least 15 time units and
make a linear fit in the range $t\in [5,15]$.

\section*{References}
\bibliographystyle{iopart-num}
\bibliography{integral_term}

\end{document}